\documentclass[10pt,letterpaper,twocolumn,prl,showpacs]{revtex4-1}
\usepackage{graphicx,color,textcomp}
\usepackage{mathrsfs,amsmath}   %The amsmath package is included for \xrightarrow
\usepackage{psfrag,overpic,pst-all}
\usepackage{amssymb}

\begin{document}

\title{High-energy soliton fission dynamics in multimode GRIN fiber}

\author{Mario Zitelli$^{1}$}
\author{Fabio Mangini$^{2}$}
\author{Mario Ferraro$^{1}$}
\author{Alioune Niang$^{2}$}
\author{Denis Kharenko$^{3,4}$}
\author{Stefan Wabnitz$^{1,3}$}

\affiliation{$^1$Department of Information Engineering, Electronics and Telecommunications (DIET), Sapienza University of Rome, Via Eudossiana 18, 00184 Rome, Italy}
\affiliation{$^2$Department of Information Engineering (DII), University of Brescia, Via Branze 38, 25123 Brescia, Italy}
\affiliation{$^3$Novosibirsk State University, Pirogova 1, Novosibirsk 630090, Russia}
\affiliation{$^4$Institute of Automation and Electrometry SB RAS, 1 ac. Koptyug ave., Novosibirsk 630090, Russia}

\begin{abstract}
The process of high-energy soliton fission is experimentally and numerically investigated in a graded-index multimode fiber. Fission dynamics is analyzed by comparing numerical observations and simulations. A novel regime is observed, where solitons produced by the fission have a nearly constant Raman wavelength shift and same pulse width over a wide range of soliton energies.  
\end{abstract}

\maketitle

%%%%%%%%%%%%%%%%%%%%%%%%%%%%%%%%%%%%%%%%%%%%%%%%%%%%%%%%%%%%%%%%%%%%%%%%%%%%%%%%%%%
\section{Introduction}
The concept of optical soliton propagation in multimode (MM) fibers was introduced in the literature nearly forty years ago \cite{Hasegawa:80,Crosignani:81,Crosignani:82}. 
%However, the behavior of these fibers is not yet completely explored. 
Although Raman soliton generation in a graded-index (GRIN) MM fiber was first demonstrated back at the end of the 80's \cite{Grudinin:88}, it is not until recent years that nonlinear optics in MMFs fibers has seen a revival of interest \cite{Mecozzi:12,doi:10.1063/1.5119434}. %thanks to their potential use for next-generation telecommunications systems and high-energy fiber lasers. 

In 2013, Renninger and Wise observed that MM solitons result from the simultaneous compensation of chromatic dispersion and modal dispersion \cite{Renninger2012R31}. The associated Raman-induced MM soliton self-frequency shift (SSFS) in GRIN MM fibers was investigated in a low pulse energy (up to 3 nJ) situation, where the MM soliton is essentially carried by the fundamental mode of the MM fiber. %Optical MM soliton stability was investigated numerically in [10].
The spatiotemporal behavior of MM solitons, including their formation and fission, was later investigated by Wright et al. with input beams leading to input energy distribution among several guided modes \cite{Wright:15}; Raman MM solitons at 2100 nm could be generated from a GRIN fiber, by using 500 fs input pulses at 1550 nm, with energies up to 21 nJ. 

%In \cite{Blow:88, Gouveia-Neto:89} it was shown for the first time the suppression and manipulation of the soliton self-frequency shift (SSFS) in a SM fiber; it was demonstrated  in particular that the SSFS can be suppressed by a pump signal providing a sufficient Raman gain.

%In \cite{Betourne:09}  it was investigated the SSFS cancellation using a nonlinear propagation of sub-nanosecond pulses in a solid-core photonic bandgap filter (PBGF), by launching pulses with a few kW peak power.

%In \cite{renninger} they studied on theoretical and experimental point of view the optical solitons and soliton self-frequency shifting in GRIN MM fibers, they observed that the process  is responsible for compensation of group-velocity dispersion in time also compensates for the separation of the different spatial modes (modal dispersion).

In the regime of highly nonlinear ultrashort pulse propagation in the anomalous-dispersion regime of a GRIN MM fiber, a variety of spatio-temporal effects was observed \cite{Wright2015R31}. 
%reminiscent of nonlinear optics in bulk media, such as self-focusing and multiple filamentation, were observed at a fraction of the usual power
By adjusting the spatial initial conditions, megawatt, ultra-short pulses, tunable between 1550 and 2200 nm, could be generated from the 1550 nm pump wavelength \cite{Wright2015R31,PhysRevLett.115.223902}.
When propagating in a few-mode GRIN MM fiber, it was shown that MM solitons display a continuous range of spatiotemporal properties, that depend on their modal composition \cite{Zhu:16}. The generation of 120 fs MM solitons was reported, with higher energies with respect to single mode (SM) solitons, a property attributed to the transverse spatial degrees of freedom that result from the fiber supporting multiple spatial eigenmodes. The modal distribution of the input beam plays an important role in controlling the different frequency conversion processes that shape the output spectrum \cite{Eftekhar:17}. Namely, the interplay of soliton fission processes, stimulated Raman scattering, dispersive wave generation and intermode four-wave mixing. Whereas in the case of step-index fibers, the presence of Raman scattering may lead to novel soliton self-mode conversion effects \cite{Rishoj:19}.

The impact of self-imaging on soliton propagation in GRIN MM fibers was recently investigated using a simplified model \cite{Ahsan:18}, based on the SM generalized nonlinear Schr\"odinger equation (GNLSE) with a spatially varying effective mode area \cite{Conforti:17}, evolving according to the variational approach (VA) \cite{Karlsson:92}. The Raman-induced fission of higher-order solitons into fundamental solitons was numerically described in terms of the SM model \cite{Ahsan:19}. The experimental validation of the SM soliton fission theory and the subsequent SSFS of different Raman solitons in MM GRIN fibers has not yet been reported.

In this paper, we experimentally study the fission of high input energy (up to 550 nJ), MW peak power femtosecond high-order MM solitons in a GRIN fiber. Input multi-soliton pulses undergo Raman-induced fission into multiple fundamental MM solitons. Our experiments permit to describe the range of validity of the reduced SM soliton description. Moreover, we also reveal several unexpected properties of soliton fission in MM fibers. Specifically, we observed that, in the high input energy regime, nonlinear losses owing to side scattering in the first few centimeters of the fiber lead to output energy clamping and SSFS suppression. MM solitons spontaneously  self-organize into a wavelength multiplex, before undergoing SSFS. Moreover, the temporal duration of each MM soliton remains a constant, in spite of their different wavelengths and associated dispersions. Moreover, the analysis of the order of the generated fundamental solitons reveals their inherently MM nature.

%%%%%%%%%%%%%%%%%%%%%%%%%%%%%%%%%%%%%%%%%%%%%%%%%%%%%%%%%%%%%%%%%%%%%%%%%%%%%%%%%%%
\section{Methods}
\subsection{Experimental setup}
\label{subsec:SET}

\begin{figure}[htbp]
\centering
{\includegraphics[width=0.8\linewidth]{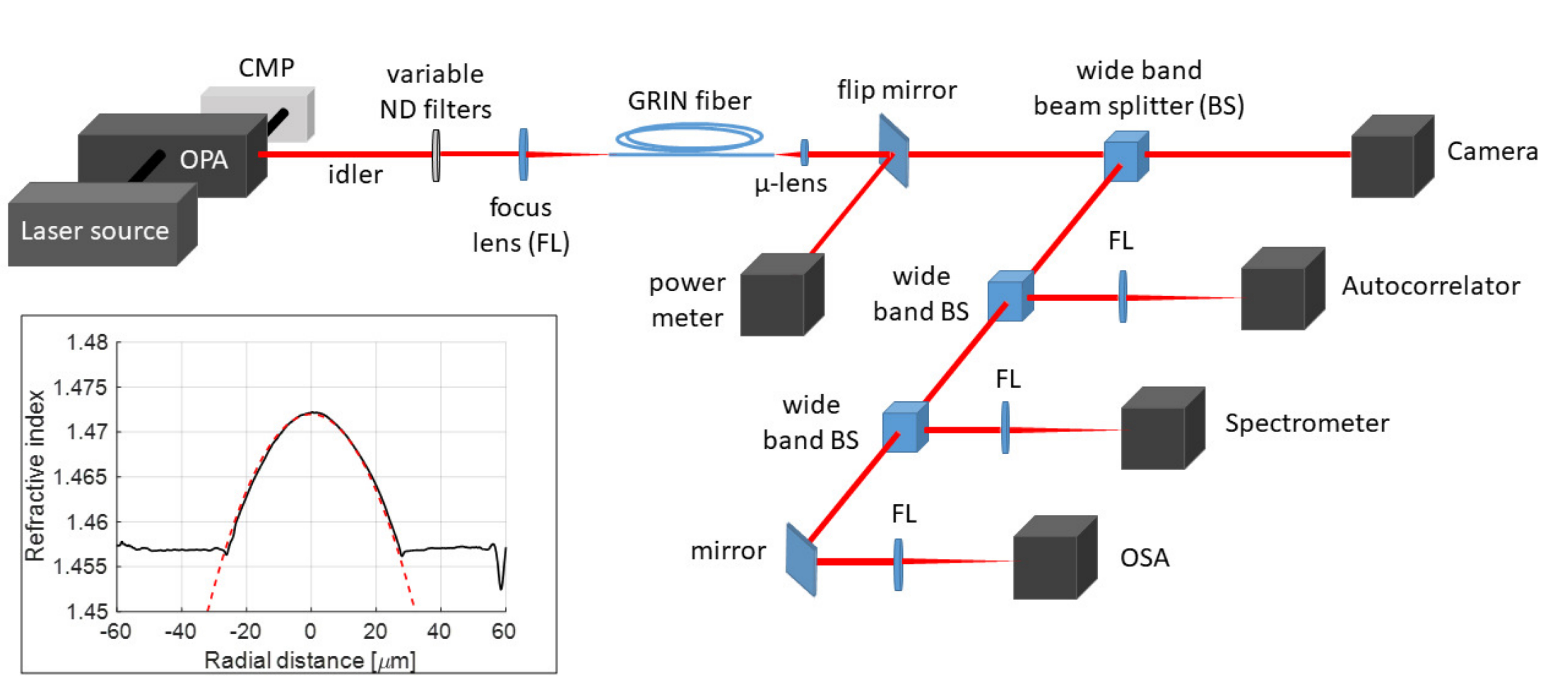}}
\caption{Schematic of the experimental setup and measured index profile of the GRIN fiber (inset).}
\label{fig:setup}
\end{figure}

As shown in Figure \ref{fig:setup}, the experimental setup used for the generation of high-energy MM solitons consists of an ultra-short laser system, involving a hybrid optical parametric amplifier (OPA) of white-light continuum (Lightconversion ORPHEUS-F), pumped by a femtosecond Yb-based laser (Lightconversion PHAROS-SP-HP), generating pulses at 25 to 100 kHz repetition rate, with Gaussian beam shape ($M^2$=1.3). The pulse shape was measured using an autocorrelator (APE PulseCheck type 2), resulting in a sech temporal shape  with 130 fs pulse width. The laser beam was focused by a 30 mm lens into a 30 cm span of GRIN fiber, with an input diameter of approximately 20 $\mu$m at $1/e^2$ of peak intensity. The input pulse wavelength is set at 1550 nm, and its energy is varied between 0.1 and 500 nJ by means of a variable external attenuator. 

The fiber index profile had a core radius $r_\text{c}$=25 $\mu$m, cladding radius 62.5 $\mu$m, and a cladding index $n_\text{clad}$=1.457, with the relative core-cladding index difference $\Delta$=0.0103. The fiber parameters were measured along with the index profile by a NR-9200 Optical Fiber Analyzer, as shown in the inset of Fig. \ref{fig:setup}.
The index measurement was based on the refracted near field. After the cleave, the fiber was immersed in a calibrated oil solution, and illuminated with a single mode laser at 667 nm. A scanning of the fiber face allowed to obtain the refractive index profile, with a resolution of 0.4 $\mu$m.
At the fiber output, a micro-lens followed by a second lens focused the beam into an optical spectrum analyzer (OSA) (Yokogawa AQ6370D) and a real-time multiple octave spectrum analyzer (Fastlite Mozza), with wavelength ranges of 600-1700 nm and 1000-5000 nm, respectively. A miniature fiber optics spectrometer (Ocean Optics USB2000+), with 200-1100 nm spectral range, was also used. The output radiation was passed through different long-pass or band-pass optical filters, in order to isolate the individual Raman solitons from the residual radiation.
Output soliton spectra were analyzed by using a $\text{sech}^2$ fit of each corresponding lobe in the output spectrum. This permitted us to determine the soliton central wavelength and bandwidth. The MM soliton energy was extracted, and compared with the total spectral power as measured by a power meter. The output soliton pulsewidth was calculated by assuming a transform limited time-bandwidth product, and checked by the autocorrelator. The output near field was also projected on a near-infrared (NIR) camera (Gentec Beamage-4M-IR), in order to optimize input coupling, and monitor the output transverse intensity distribution. Input and output average powers were measured by a thermopile power meter (GENTEC XLP12-3S-VP-INT-D0).

%%%%%%%%%%%%%%%%%%%%%%%%%%%%%%%%%%%%%%%%%%%%%%%%%%%%%%%%%%%%%%%%%%%%%%%%%%%%%%%%%%%
\subsection{Numerical model}
\label{subsec:NUM}

Our numerical model for simulating high-energy optical pulse propagation in a GRIN MMF involves the $(3D+1)$ GNLSE (or Gross-Pitaevskii equation \cite{pitaevskii}) in its vectorial form, involving a single field for each polarization, including all frequencies and modes: 
\begin{align}\label{3d+1}
&\frac{\partial A_p(x,y,z,t)}{\partial z}=\frac{i}{2k} \bigg(\frac{\partial^2A_p}{\partial x^2}+\frac{\partial^2 A_p}{\partial y^2}\bigg)\\ \nonumber
&-i\frac{\beta_2}{2}\frac{\partial^2A_p}{\partial t^2}+ \frac{\beta_3}{6}\frac{\partial^3A_p}{\partial t^3}\\ \nonumber 
&+\frac{\beta_4}{24}\frac{\partial^4A_p}{\partial t^4} -\frac{\alpha}{2}A_p+\\ \nonumber
+&i\frac{k}{2}\bigg[\frac{n^2(x,y)}{n^2_{clad}}-1\bigg]A_p+
i\gamma \bigg(1+iK_2+\frac{i}{\omega_0}\frac{\partial}{\partial t}\bigg)\\ \nonumber
&\bigg[(1-f_R)A_p\bigg(|A_p|^2+\frac{2}{3}|A_q|^2+\frac{1}{3}A_p^2e^{-2i\omega_0t}\bigg)+\\ \nonumber
+& f_RA_p\int_{-\infty}^t d\tau h_R(\tau)\bigg(|A_p(t-\tau)|^2+\frac{2}{3}|A_q(t-\tau)|^2\bigg)\bigg]
\end{align}
with $\gamma=n_2\frac{2\pi}{\lambda}$, $K_2=\frac{\alpha_2}{2\gamma}$, and $f_R=0.18$.
In Eqs.(\ref{3d+1}), the two polarizations $p,q=x,y$ are nonlinearly coupled. Terms in the right-hand side of Eq. (\ref{3d+1}) account for: transverse diffraction, second, third and fourth-order dispersion, linear loss, the waveguiding term with refractive index profile $n(x,y)$ and cladding index $n_\text{clad}$, Kerr and Raman nonlinearities (with nonlinear coefficient $\gamma$  and fraction  $f_R$), respectively. In Eqs.(\ref{3d+1}) we neglect the contribution of polarization mode dispersion: we numerically verified that its effects are negligible for the short fiber lengths (up to 30 centimeters) involved in our experiments. Nonlinearities include self-steepening, third-harmonic generation (THG) and two-photon absorption (TPA) (with coefficient $\alpha_2$). The model of Eq. (\ref{3d+1}) is alternative to that based on coupled equations for the fiber modes \cite{Poletti:08}, which requires a preliminary knowledge of the power distribution among the fiber modes at the fiber input.
Random phase noise was added to the input field, in order to describe the generation of intensity speckles and to seed dispersive waves, which are typically observed in multimode fibers. Random changes in the fiber core diameter have also been introduced, although they produced negligible effects over the considered fiber lengths.
%; polarization mode dispersion was also neglected.

%$r_c=25$ $\mu$m, cladding radius 62.5 $µ$m, cladding index $n_{clad}=1.457$, relative index difference $\Delta=0.0103$ (the index profile parameters are tested experimentally)
In simulations, we used the following GRIN fiber parameters: core radius, cladding radius, and relative index difference are already provided in the experimental setup subsection; dispersion parameters are $\beta_2=-22\, \text{ps}^2/\text{km}$ at 1550 nm, $\beta_3=0.132\,\text{ps}^3/\text{km}$, $\beta_4=-5\times 10^{-4}\, \text{ps}^4/\text{km}$; nonlinear parameters are $n_2=2.7\times10^{-20}\, \text{m}^2/\text{W}$, $\alpha_2=1\times10^{-16}\, \text{m}/\text{W}$ (this value was justified by a best fit of the SSFS observed in the experiments); $h_R(\tau)$ with the typical response times of 12.2 and 32 fs, respectively \cite{Stolen:89,Agrawal01}. We included the wavelength dependence of the linear loss coefficient $\alpha$, as reported for standard SM glass fibers \cite{shubert}. For a 30 cm fiber length, linear losses only affect transmission at wavelengths above 2500 nm, and are negligible at the wavelengths considered in this work.
The input beam is modeled as a Gaussian beam with $w_0=10$ $\mu$m waist (20 $\mu$m diameter); we used a sech temporal shape with full-width-at-half-maximum (FWHM) pulsewidth $T_\text{FWHM}=130$ fs; random phase noise is added at input for both polarizations. Two polarizations are launched at the fiber input with a variable spit ratio between 0.5 and 1, the last value corresponding to a single polarization.
Although the used model considers all modes merged in a single field, our fiber launch conditions produce, in linear regime, 4 axial-symmetric modes with power fraction: 0.56, 0.26, 0.12, and 0.06, respectively; higher-order modes can be neglected. In the next Section \ref{sec:RES}, we shall compare numerical simulations using Eqs.(\ref{3d+1}) to experimental results. 
%%%%%%%%%%%%%%%%%%%%%%%%%%%%%%%%%%%%%%%%%%%%%%%%%%%%%%%%%%%%%%%%%%%%%%%%%%%%%%%%%%%
\section{Results and discussion}
\label{sec:RES}

Here we report our experimental studies of MM soliton fission dynamics, using the setup of Subsection \ref{subsec:SET}. We considered relatively short GRIN fiber spans of length up to 30 cm. For a better understanding of the complex MM nonlinear effects involved, we compared observations with numerical simulations with the full MM model described in Subsection \ref{subsec:NUM}, as well as with the analytical theory for SM Raman solitons \cite{Gordon:86}. All figure captions specify if the curves refer to either experimental or simulation data.

 \subsection{Nonlinear loss}
\label{subsec:NLOSS}

\begin{figure}[htbp]
\centering
{\includegraphics[width=0.475\linewidth]{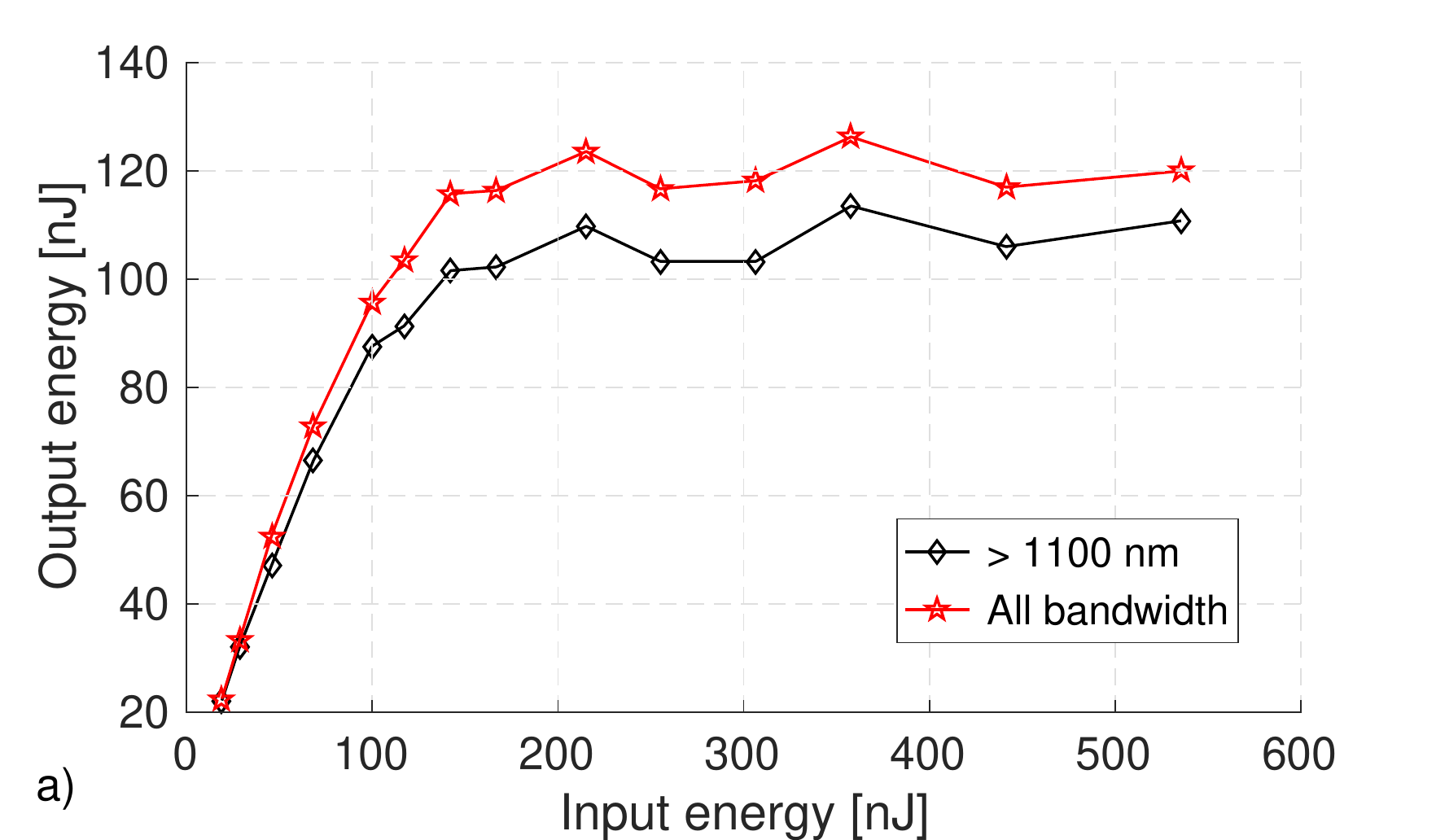}}
{\includegraphics[width=0.475\linewidth]{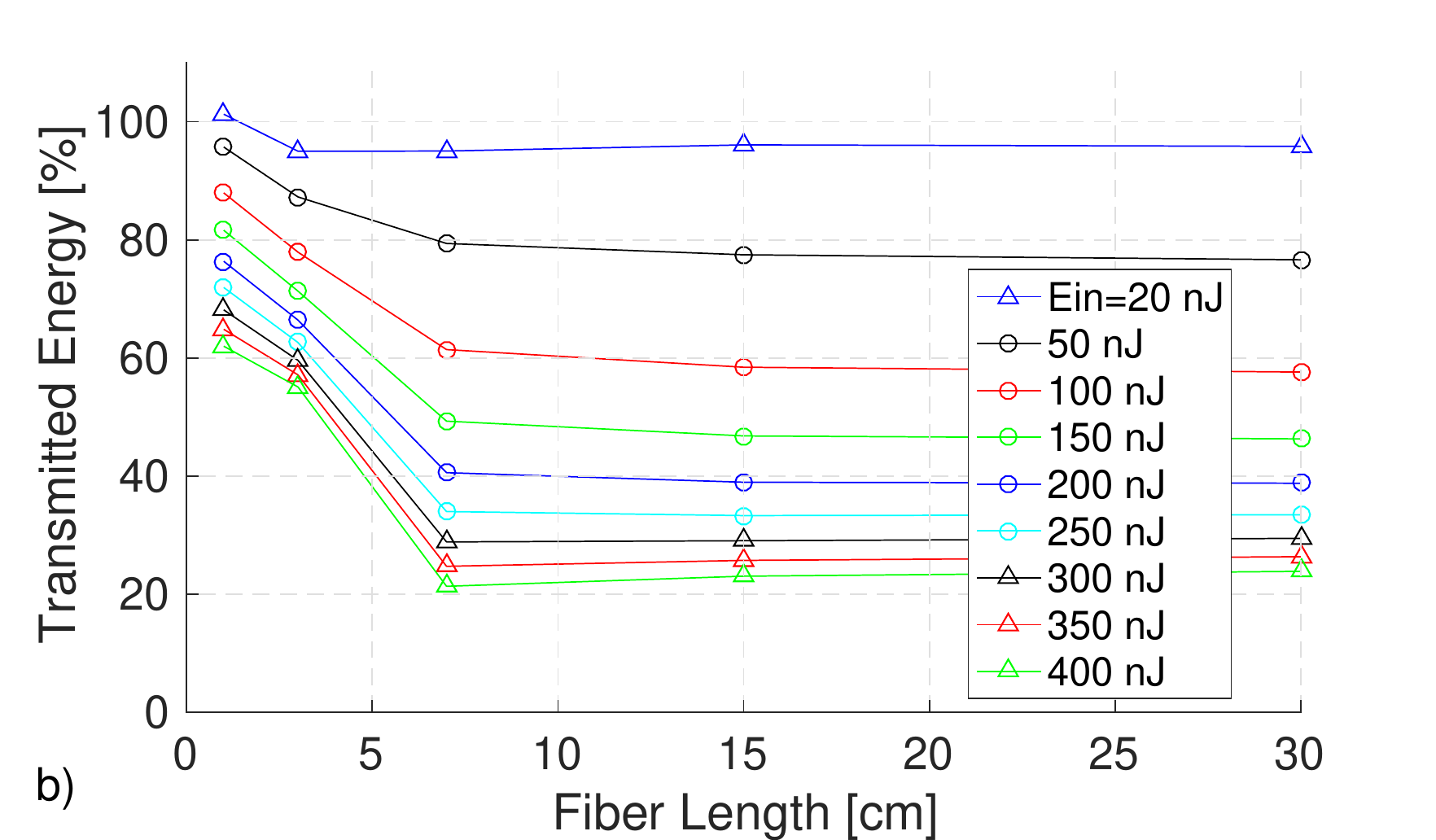}}
\caption{a) Experimental output pulse energy vs. input energy, in the overall bandwidth or for wavelength above 1100 nm. b) Fraction of transmitted energy vs. fiber length for different input pulse energies.}
\label{fig:Trans}
\end{figure}

 We observed a strong output energy clamping for an input pulse energy $E_\text{in}$>100 nJ. 
 Figure \ref{fig:Trans}.(a) shows the measured output energy vs. input energy, in the overall, detected spectral bandwidth (red curve with stars), or after insertion of a filter that strongly attenuates all wavelengths below 1100 nm (black curve with diamonds). In both cases, energy losses are limited from zero to below 20\% for input energies $E_\text{in}$ up to 100 nJ. For $E_\text{in}>$150 nJ, the output energy saturates to a nearly constant value. This output energy clamping is not affected by changing the laser pulse repetition rate, while maintaining the input pulse energy unchanged, indicating that thermal effects are not involved in the output energy loss.
 
Figure \ref{fig:Trans}.(b) is the result of a cutback experiment (performed with a 50 mm input lens), illustrating the dependence of nonlinear energy loss upon fiber length (starting from 1 cm), for several values of the input pulse energy. Here the loss is determined in terms of the transmitted energy by the fiber or the ratio of output to input pulse energy. This ratio is normalized to the fiber coupling efficiency at low powers (equal to 45\%), so that the transmitted energy at low powers is 100\% for all fiber lengths. Figure \ref{fig:Trans}.(b) shows that losses mainly occur over the first 7 cm of fiber (and particularly over the first cm already), and remain negligible in the remaining fiber section. The strong drop of transmitted energy as the input energy grows larger confirms the nonlinear nature of the loss mechanism.
We ascribe the nonlinear loss to multi-photon absorption, leading to the side-emission of broadband blue fluorescence \cite{doi:10.1063/1.5119434}, as well as to scattering from the cladding of NIR radiation, THG, and visible dispersive wave sidebands \cite{Wright2015R31,PhysRevLett.115.223902} . %a second mechanism, less important, is two-photon absorption. Both mechanism.
Nonlinear loss is most efficient at the points of peak beam intensity generated by self-imaging over the first few cm of fiber. A maximum is reached during the input MM soliton pulse fission phase, occurring in the first centimeters of fiber, were the peak power reaches several MWs \cite{doi:10.1063/1.5119434,Conforti:17}. We defer a detailed comparison between experiments and theory of multi-photon absorption and side-scattering induced nonlinear loss processes in the GRIN fiber to a subsequent publication.  

Two energy regimes are observed in Figure \ref{fig:Trans}.(a) and \ref{fig:Trans}.(b). In the first regime, for $E_\text{in}$<100 nJ, nonlinear losses are negligible. Whereas for 100<$E_\text{in}$<550 nJ, significant nonlinear loss occurs over the first few cm of fiber, and the output energy is clamped. In the following, we will refer to the first and second regime as ``low-loss'' and ``high-loss'' cases, respectively. We also observed that, for input energies above 550 nJ (i.e., about 2.1 MW input peak power), the beam undergoes catastrophic collapse with permanent damages at the input and across the fiber length.

%(see Fig\ref{fig:microscope}).
%\begin{figure}[htbp]
%\centering
%\fbox{\includegraphics[width=0.75\linewidth]{foto_fibra}}
%\caption{Microscopic image of the first 6 mm of GRIN fiber acquired by %Dinocapture-ligth2.0 digital microscope.}
%\label{fig:microscope}
%\end{figure}

%We measured by a cutback experiment the energy loss vs. fiber length at several input pulse energies. Loss was determined in terms of the transmitted energy by the fiber, or the ratio of output to input pulse energy, normalized to the fiber coupling efficiency at low powers (the transmitted energy at low powers is 100\% for all fiber lengths). Our analysis shows that energy losses largely occur over the first 7 cm of fiber, and remain negligible in the remaining section, which confirms the nonlinear nature of the losses.

 \subsection{Fission dynamics}
\label{subsec:FISS}

In order to analyze the MM soliton fission process and compare it with existing SM soliton theory predictions, we carried out a series of simulations in the ``low-loss'' regime, where the impact of TPA can be neglected. 

\begin{figure}[htbp]
\centering
{\includegraphics[width=0.475\linewidth]{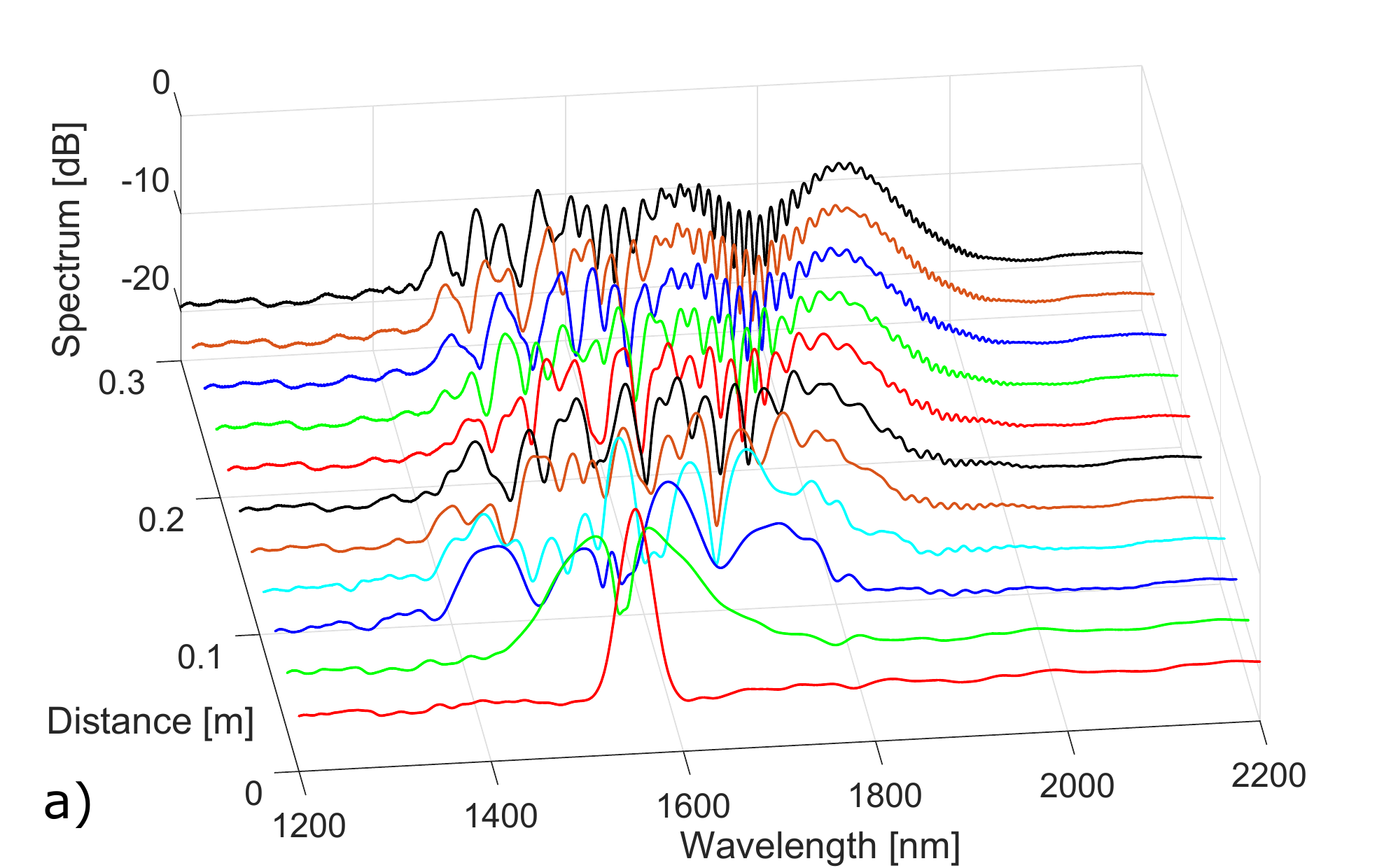}}
{\includegraphics[width=0.475\linewidth]{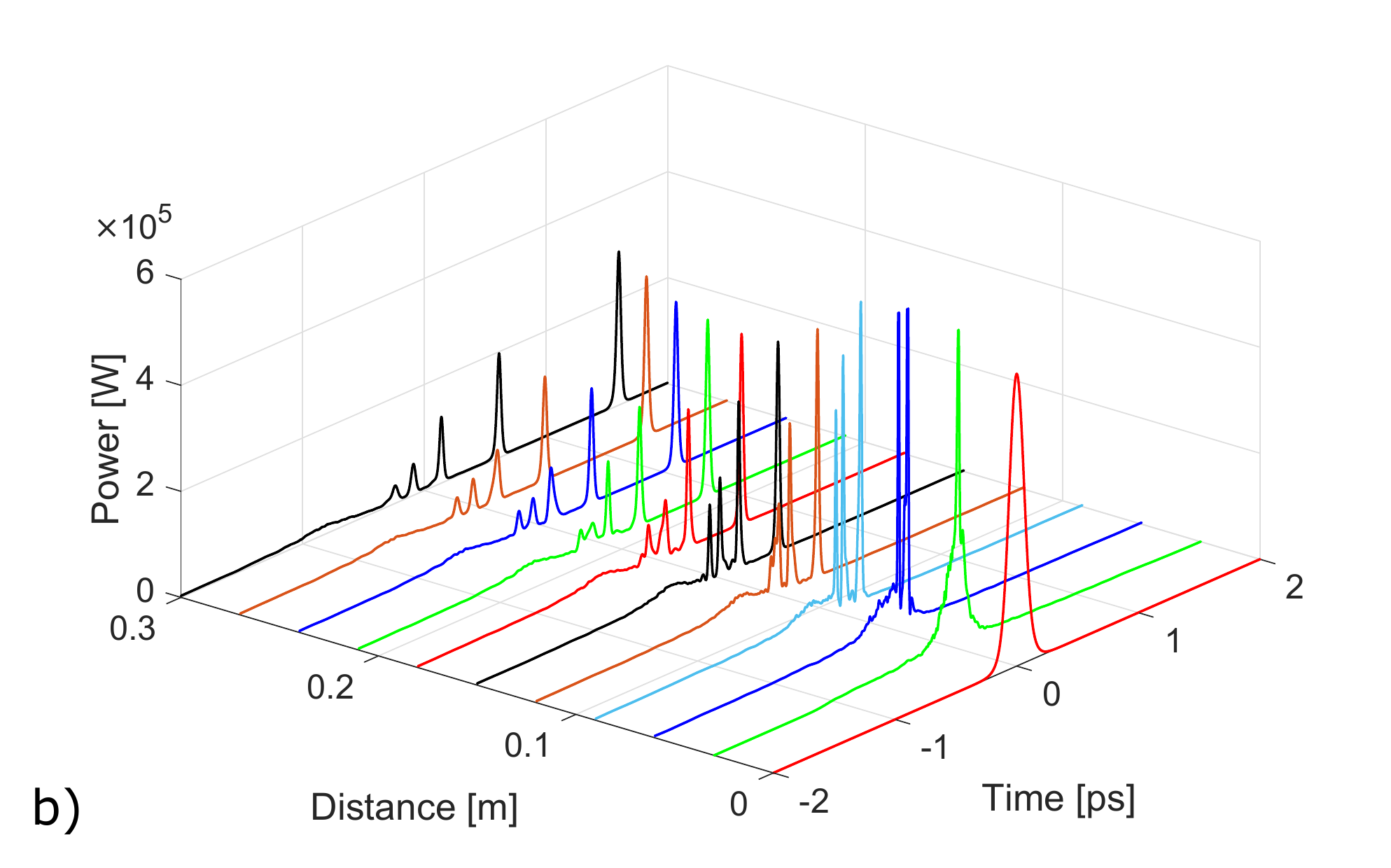}}
{\includegraphics[width=0.475\linewidth]{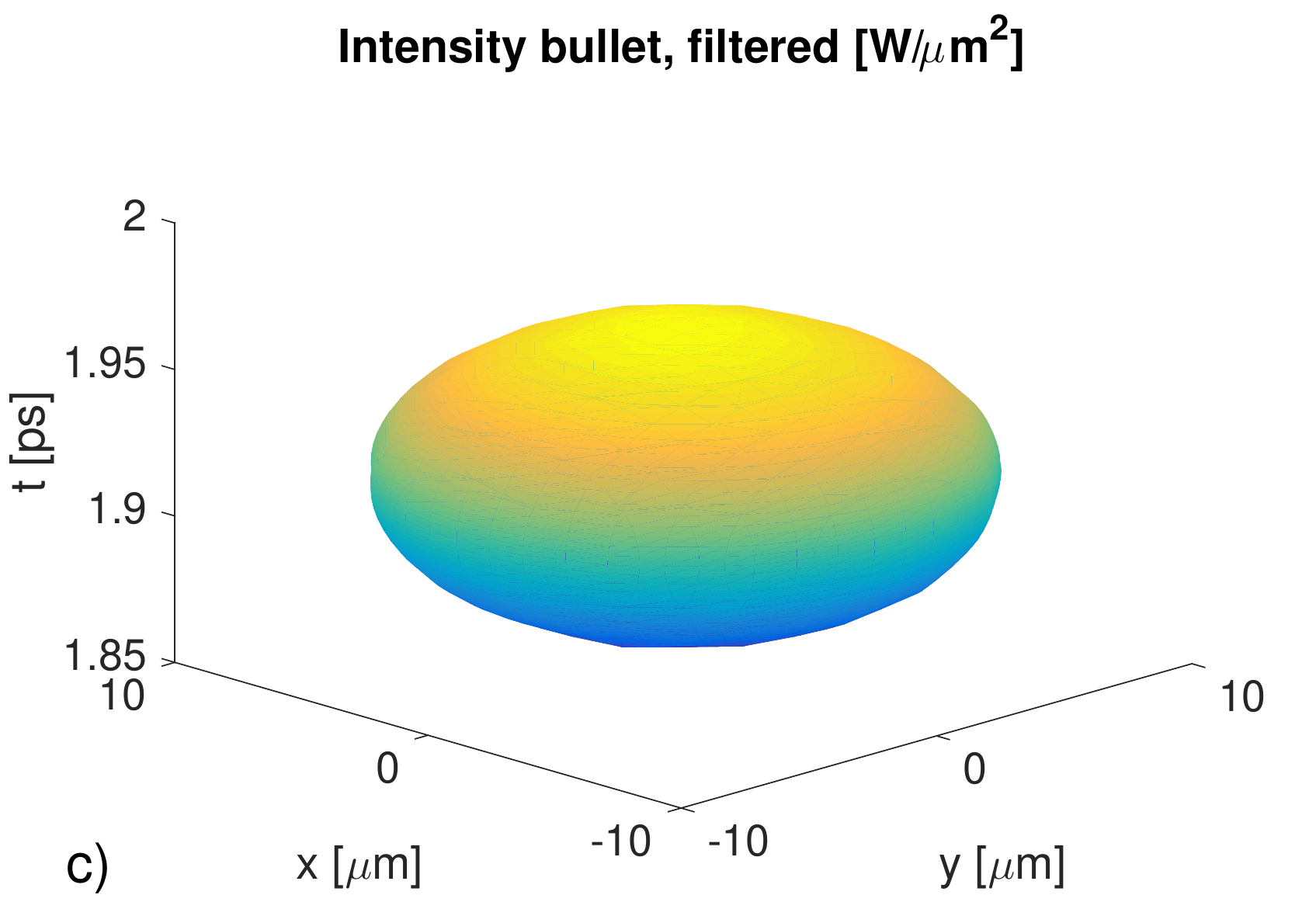}}
{\includegraphics[width=0.475\linewidth]{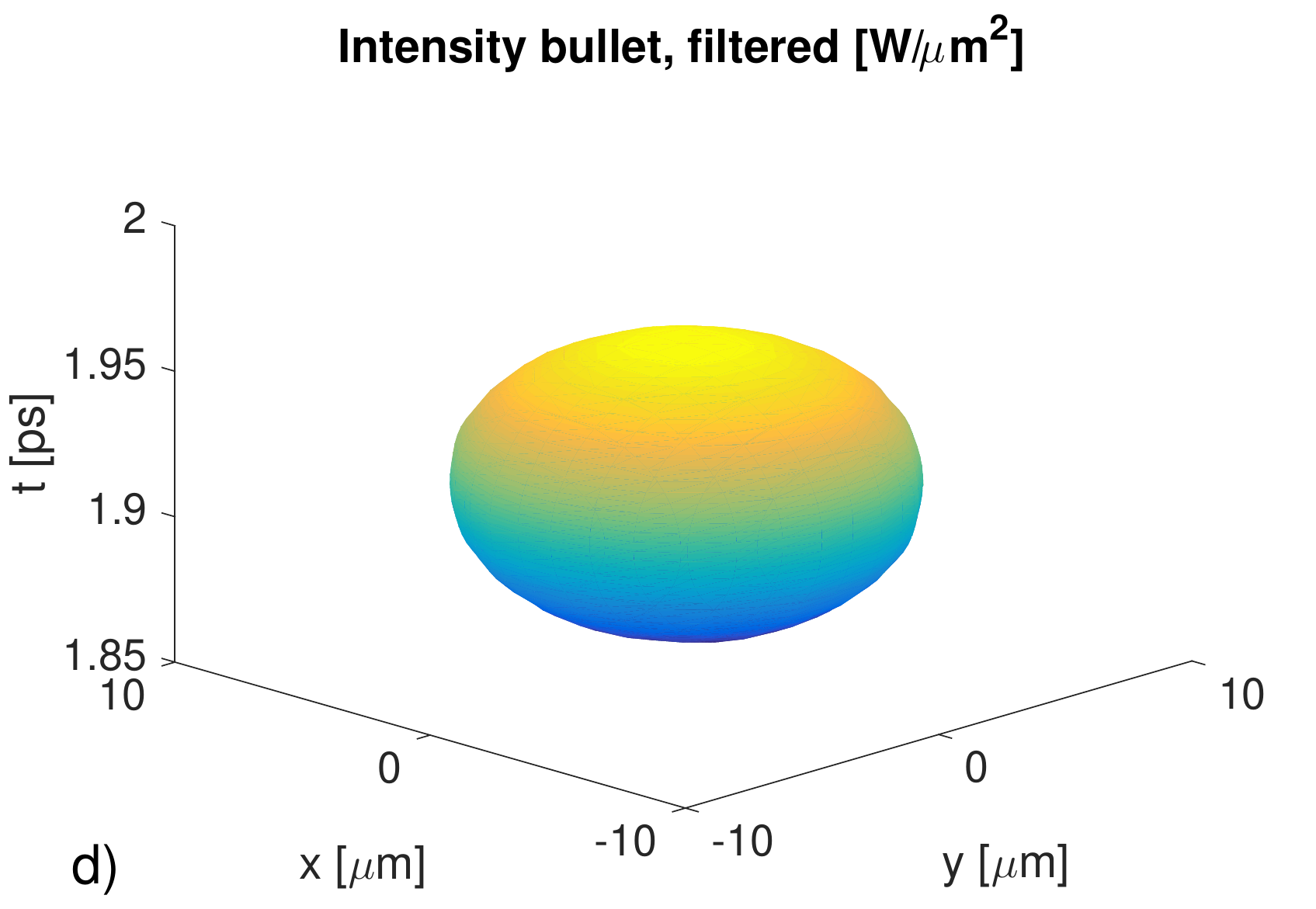}}
\caption{Simulation of the a) spectra, and b) pulse power evolution with distance; case of 90 nJ input pulse energy. c) Simulation of the most delayed bullet in space-time domain; last self-imaging period before the output, at the distance of maximum effective area; d) last self-imaging period, at the distance of minimum effective area. Iso-surfaces are provided at $1/e^2$ of peak intensity.}
\label{fig:Simulations}
\end{figure}

%Figures \ref{fig:Simulations}.a and \ref{fig:Simulations}.b  are simulations that better illustrate the fission process occurring in the fiber. 
Fig. \ref{fig:Simulations}.(a) shows the evolution of the intensity spectrum as a function of fiber length, for the input energy of 90 nJ.
As can be seen, the fission of the input pulse leads to a complex wavelength multiplex of MM solitons, which is generated at the short distance of about ten centimeters. The resulting soliton wavelength shifts are nonadiabatic, i.e., they occur over a much shorter distance than the length required to obtain an equivalent wavelength shift $\Delta \lambda$ via the SM SSFS formula \cite{Gordon:86}

\begin{equation}\label{lambdaa}
  \Delta\lambda=\frac{4\lambda^2\beta_2T_Rz}{15\pi cT_0^4}
\end{equation}

\noindent where $T_R$=3 fs is the Raman response time, $z$ is the fiber propagation coordinate, $T_0$=$T_\text{FWHM}/1.763$ is the soliton pulse width, and $\lambda$ is the soliton wavelength.
After the fission process, each resulting MM soliton in the wavelength multiplex undergoes a slow propagation frequency shift, which may be well described according to Eq.~\eqref{lambdaa}. 

When displaying the pulse power in the time domain, i.e., by integrating the field intensity across the transverse $x$, $y$ axes, the simulation of Fig. \ref{fig:Simulations}.(b) shows that the input pulse undergoes a temporal compression over the first 5 mm of fiber. Correspondingly, it reaches local peak powers up to several MW until it experiences a soliton fission, thus producing a number of solitons, each with a comparable pulse width between 45 fs and 60 fs, and different peak powers, ranging from few tens of kW up to 300 kW. The number $M$ of generated solitons may arrive up to 5 when the input energy grows larger. In the ``low-loss''regime, the numerically generated solitons have energies below 20 nJ (peak powers below 300 kW).

In the spatiotemporal domain, the train of solitons generated after the input pulses fission corresponds to as many bullets. Figs. \ref{fig:Simulations}.(c) and  \ref{fig:Simulations}.(d) reports the most delayed Raman soliton bullet during the last self-imaging period, before the fiber end; bullets are illustrated as iso-surfaces at $1/e^2$ of global peak intensity, and are given at the distance of maximum (Fig. \ref{fig:Simulations}.(c)) and minimum (Fig. \ref{fig:Simulations}.(d)) effective area. Soliton bullets experience a synchronous periodic self-imaging (SI) in the transverse plane, with a period that corresponds remarkably well to the theoretical value, namely $z_\text{SI}=\pi r_c/\sqrt{2\Delta}=0.55$ mm. All soliton bullets, after integration in time, possess similar minimum and maximum effective areas, despite of their different wavelengths; the ratio of minimum to maximum effective area for all bullets is $C=\mathcal{A}_\text{min}/\mathcal{A}_\text{max}=0.53$,in good agreement with the prediction of the VA
\cite{Karlsson:92,Conforti:17,Ahsan:18}.

Fig. \ref{fig:Simulations}.(b) also shows the generation of dispersive waves with a negative delay, corresponding to anti-Stokes radiation. Our simulations show that those waves partially propagate in the cladding.

 \subsection{Soliton spectra}
\label{subsec:SPEC}

\begin{figure}[htbp]
\centering
{\includegraphics[width=0.48\linewidth]{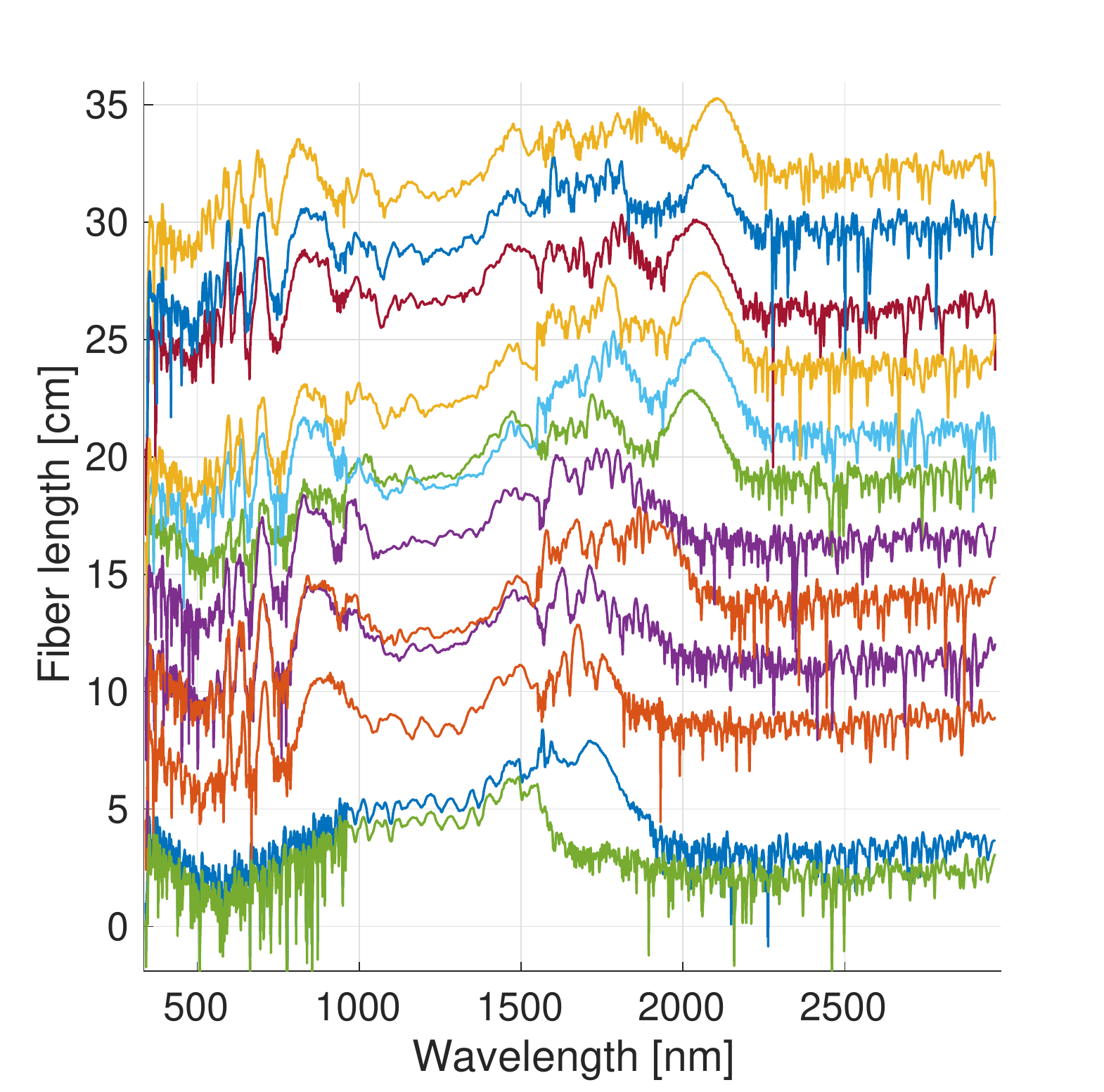}}
%{\includegraphics[width=0.475\linewidth]{Fig3.7_cutback_dB_2ver}}
{\includegraphics[width=0.48\linewidth]{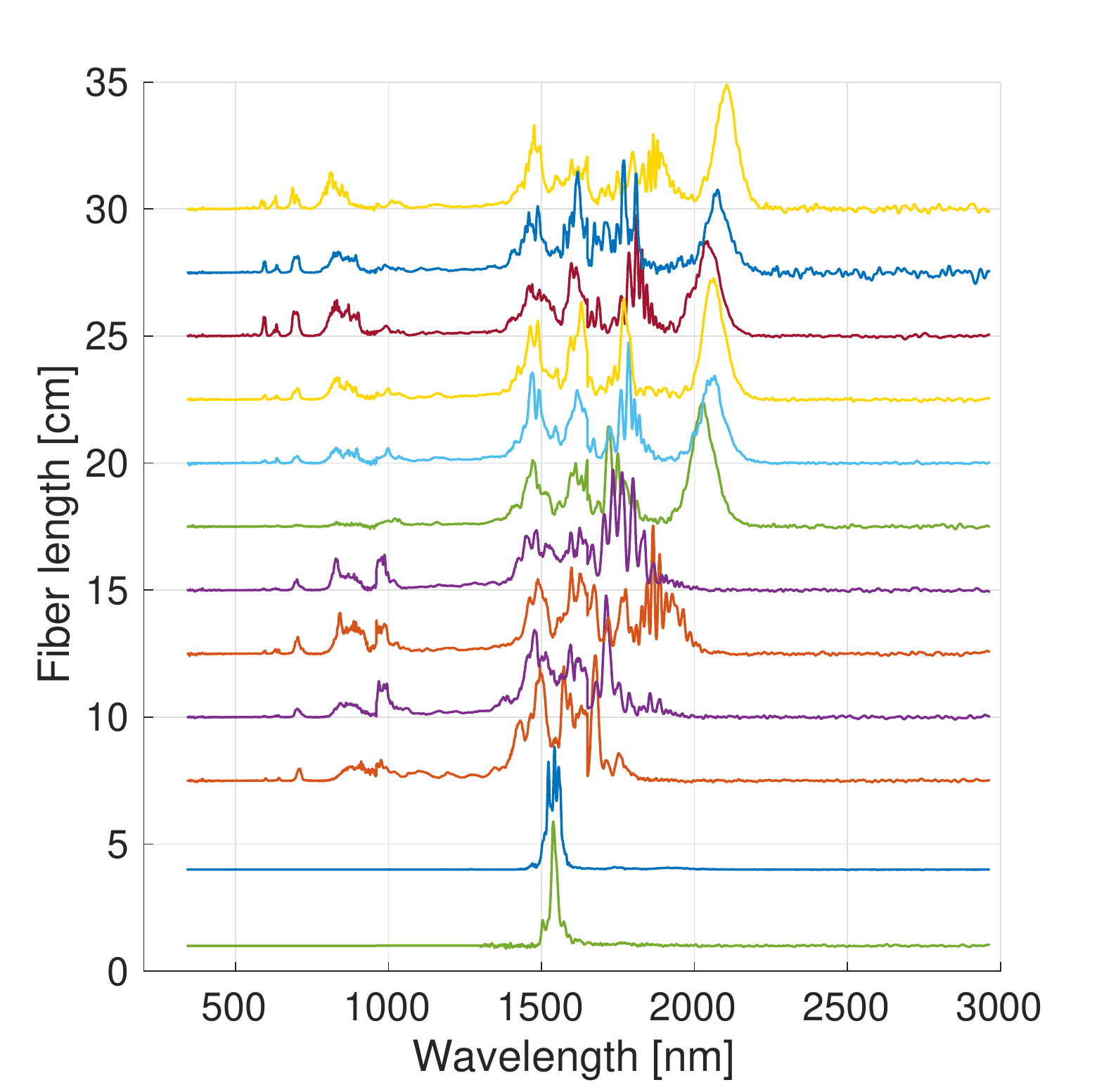}}
\caption{Cutback experiment at 300 nJ input energy. Experimental output intensity spectra in log scale (left) and in linear scale (right)}
\label{fig:Csoliton}
\end{figure}

In this Subsection, we compare the predictions of Subsection \ref{subsec:FISS} to the experiments. Moreover, we experimentally investigate the transition from the ``low-loss'' to the ``high-loss'' propagation regime, and its impact on MM soliton fission, followed by the SSFS of fundamental MM solitons. 

%Figure \ref{fig:m2camera} show the beam shape at the output of the GRIN fiber in near field for several fiber length from 7.5 to 30 cm at 2.5 cm stages.
To disclose the details of high-order MM soliton fission, and the subsequent SSFS dynamics, we carried out a cutback experiment. Figure \ref{fig:Csoliton} illustrates the length dependence of the output spectrum, for $E_\text{in}$=300 nJ. A wavelength multiplex of fundamental Raman MM solitons appears at 7 cm, as an outcome of high-order MM soliton fission. In the first 3 cm of fiber, no Raman MM solitons are visible, since the fission process has not yet been completed. The generation of anti-Stokes dispersive wave sideband series (originating from spatiotemporal MM soliton oscillations \cite{Wright2015R31,PhysRevLett.115.223902}) is visible for wavelengths below 900 nm. A close inspection of the spectra shows that visible sidebands experience a progressive, weak blue-shift as the distance grows larger.

%\begin{figure}[htbp]
%\centering
%{\includegraphics[width=0.7\linewidth]{beam_shape}}
%\caption{Beam shape in near field acquired by Gentec M$^2$ camera, during the cutback experiment at 300 nJ input energy.}
%\label{fig:m2camera}
%\end{figure}

\begin{figure}[htbp]
\centering
{\includegraphics[width=0.75\linewidth]{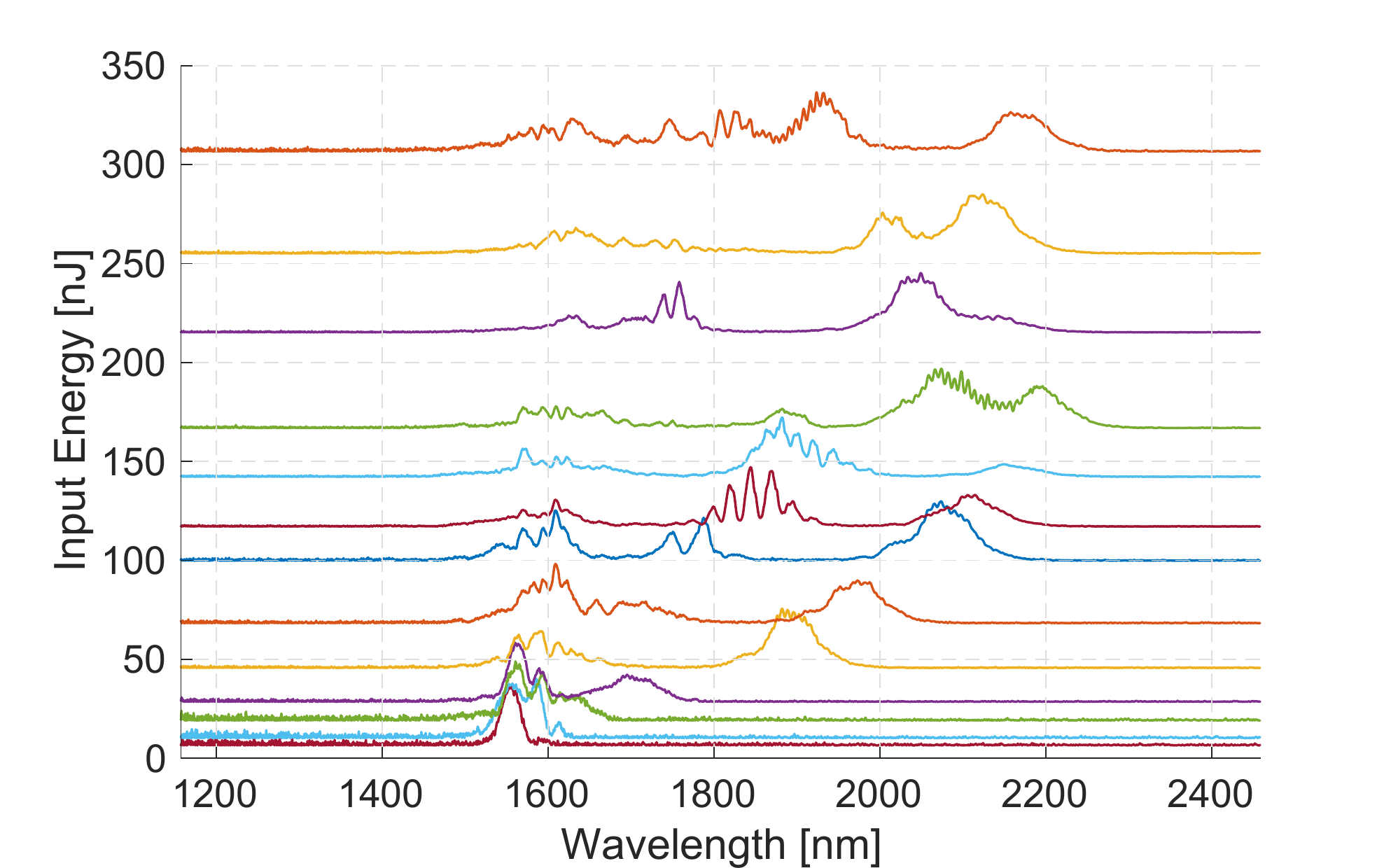}}
\caption{Experimental output spectra in linear scale, from a GRIN fiber of 30 cm length, for several values of the input energy. Points where spectra cross the vertical axis indicate the corresponding input pulse energy.}
\label{fig:SPsoliton}
\end{figure}

Figure \ref{fig:SPsoliton}, showing the experimental output spectra in a linear scale (with a GRIN fiber length of 30 cm) for several values of the input energy, illustrates well the transition from the ``low-loss'' to the ``high-loss'' regimes.
%Points where spectra cross the vertical axis indicate the corresponding input energy. 
As can be seen, a progressively larger number of MM Raman solitons is generated as $E_\text{in}$ grows larger. Individual MM solitons undergo different amounts of SSFS. In the ``high-loss'' propagation regime, for $E_\text{in}$>150 nJ, the Raman-induced wavelength shift stops at 2250-2300 nm, owing to nonlinear losses that clamp the output energy.

 \subsection{Soliton wavelengths}
\label{subsec:WAVE}

\begin{figure}[htbp]
\centering
{\includegraphics[width=0.7\linewidth]{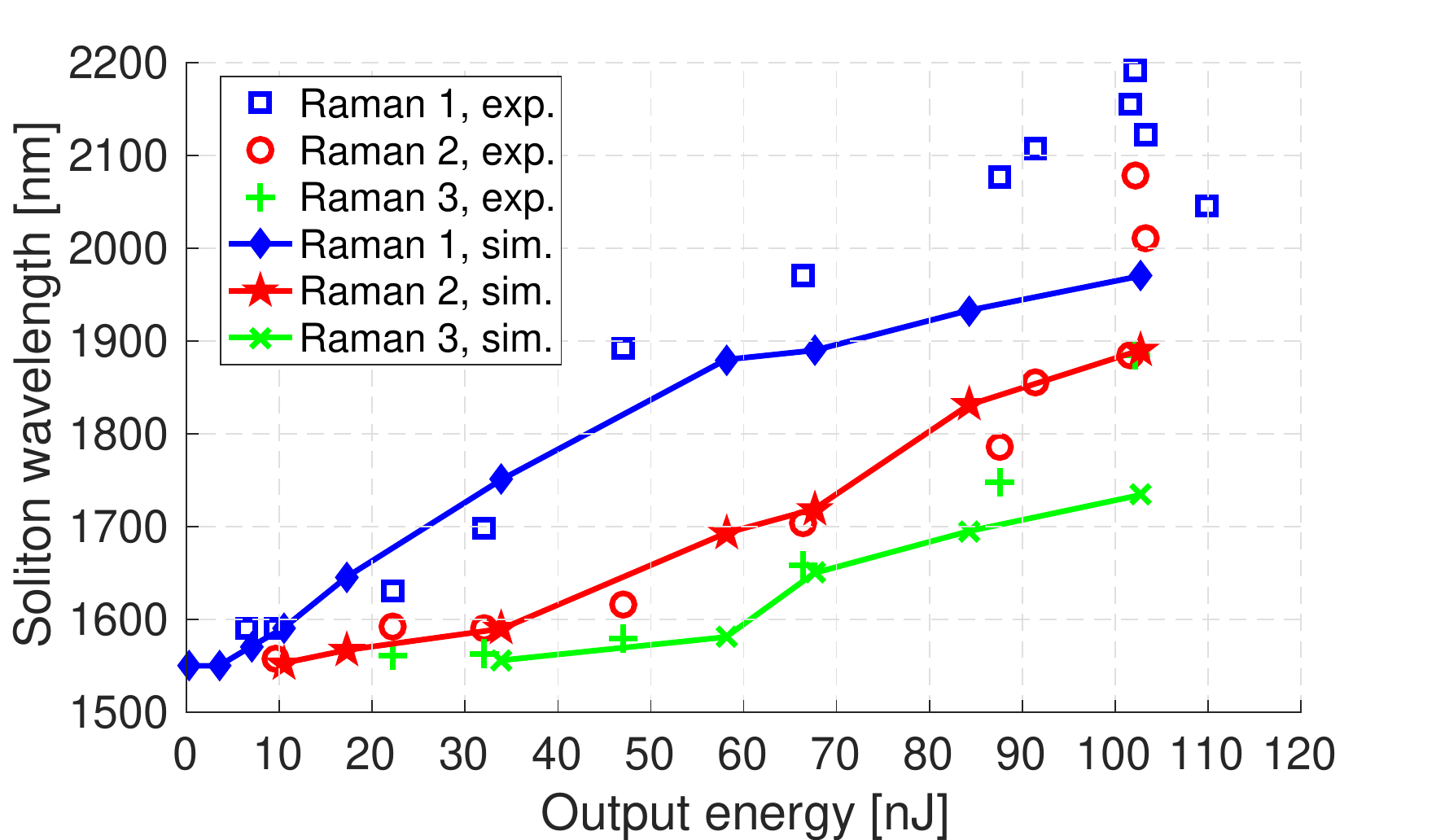}}
\caption{Experimental wavelength shifts (empty squares, empty circles and crosses for solitons Raman 1, 2 and 3, respectively) vs. total output energy, after long-pass optical filtering at 1100 nm, at 30 cm of GRIN fiber. The wavelengths of the three Raman output solitons are ranged according to their absolute wavelength shift. Corresponding simulation results are indicated by solid lines with diamonds, stars and x-shaped crosses, respectively.}
\label{fig:Ssoliton}
\end{figure}

We extracted in Figure \ref{fig:Ssoliton} the measured wavelength of individual MM solitons that are generated by the higher-order soliton fission process.

In Figure \ref{fig:Ssoliton}, we display the soliton wavelength as a function of the total output energy above 1100 nm (measured after inserting a long-pass filter, in order to eliminate most of the generated THG and supercontinuum, while including all solitons). Data in Figure \ref{fig:Ssoliton} include MM solitons generated in both the low-loss and in the high-loss regimes.

In Figure \ref{fig:Ssoliton}, we refer to Raman 1, Raman 2 and Raman 3, as the solitons that experience the highest to the lowest Raman shift, respectively. Additional solitons, if present, are not displayed here. The experimental data, illustrated as dots, are provided where the output spectra provide lobes that are easily fittable by using a $\text{sech}^2$ shape; other spectra have been discarded. Solid curves are the corresponding data from numerical simulations. In Fig.\ref{fig:Ssoliton}, numerical curves stop right above 100 nJ of output energy. As a matter of fact, simulations diverge for output energies above 110 nJ; numerics blow up due to beam filamentation after a few centimeters of fiber. As discussed in Subsection \ref{subsec:NLOSS}, in experiments the fiber damage threshold only occurs for relatively high input energies, above 550 nJ. We ascribe the discrepancy between simulations and experiments in the threshold for catastrophic self-focusing to the presence of a significant additional loss mechanism via the side-scattering of light, a mechanism which is not included in our model.

%, showing a remarkable agreement. 
%Our numerical model used the (3D+1) GNLSE (or Gross-Pitaevskii equation \cite{doi:10.1063/1.5119434}) in vectorial form, propagating a single field for each polarization.

Experimental data for the Raman 1, Raman 2, and Raman 3 solitons show that, for output energies up to 90 nJ (``low-loss'' regime), their wavelengths increase linearly with energy: the most shifted Raman 1 soliton reaches 2100 nm. For input energies above 90 nJ, high nonlinear losses provide a nearly constant output energy: as a result, an irregular distribution of soliton wavelengths appears in Figure \ref{fig:Ssoliton}, and the output energy is clamped to values below 110 nJ. 

\begin{figure}[htbp]
\centering
{\includegraphics[width=0.7\linewidth]{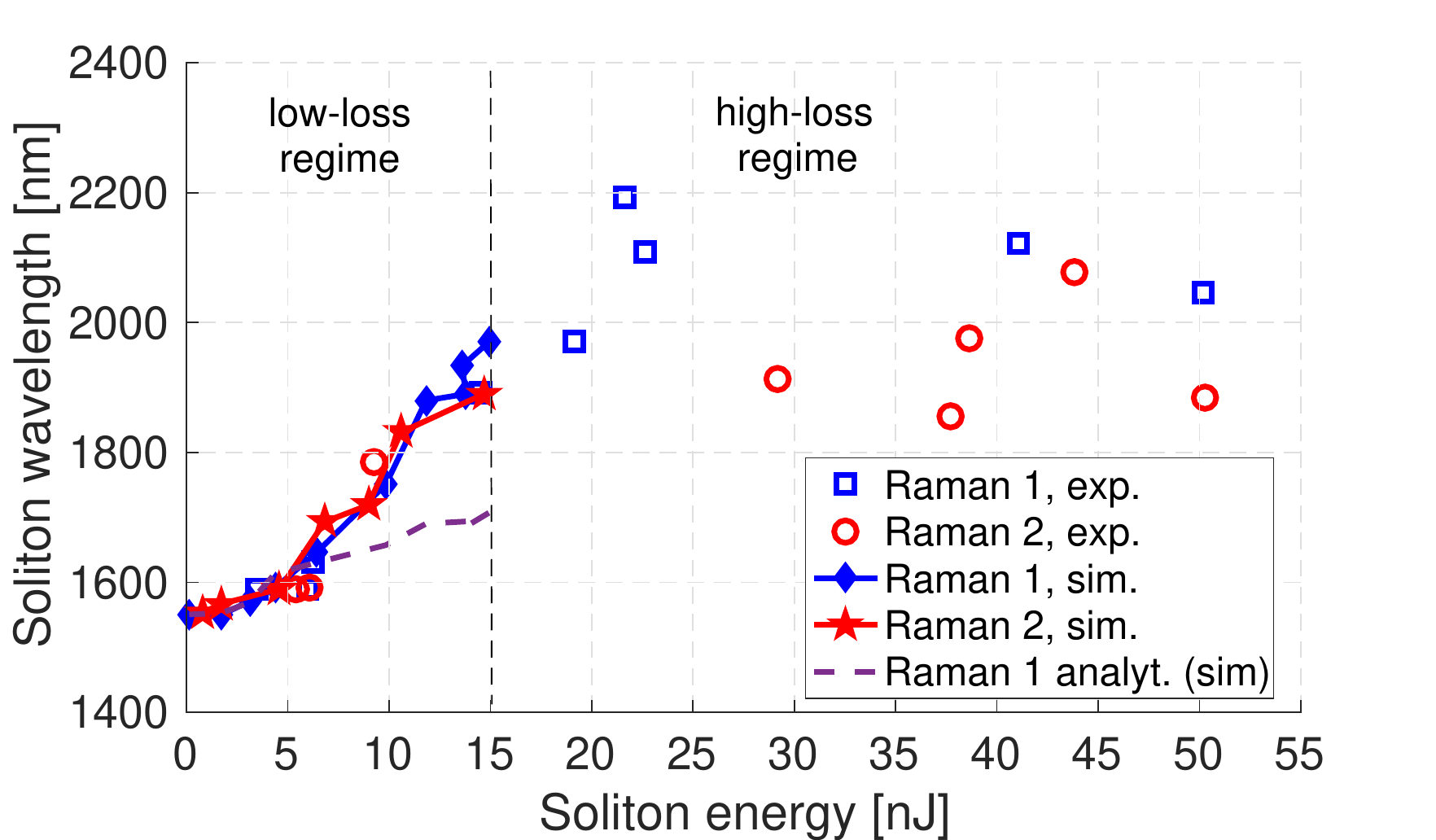}}
\caption{Experimental data (empty squares and empty circles for solitons Raman 1 and 2, respectively) and numerical simulations (solid curves) of the wavelength of solitons Raman 1 and Raman 2 vs. soliton energy; violet dashed line is obtained from Eq.~\eqref{lambdaa}.}
\label{fig:Lsoliton}
\end{figure}

%Fig. \ref{severallens} shows the wavelength of the most Raman shifted MM soliton as a function of the total energy of the input pulse, for different input beam diameters. As can be seen, the largest input beam diameter, i.e., the highest input multimode excitation leads to the smallest SSFS value.

%\begin{figure}[htbp]
%\centering
%{\includegraphics[width=0.7\linewidth]{Raman1Shift_Psolitone_fitt}}
%\caption{Wavelength of MM Raman solitons, for 18, 30, 45 mm input beam diameters, as a function of input energy.}
%\label{fig:severallens}
%\end{figure}

The difference between the ``low-loss'' and the ``high-loss'' regimes is also illustrated by Figure \ref{fig:Lsoliton}, providing experimental data (empty squares and empty circles) and numerical simulations (solid curves) of the wavelength of solitons Raman 1 and Raman 2, respectively, as a function of their output energy, $E{s}$. Here the violet dashed line is analytically obtained from the SM SSFS equation ~\eqref{lambdaa}; we inserted the pulsewidth and wavelength $\lambda$ of the corresponding output soliton from simulations; dispersion $\beta_2$ is calculated at soliton wavelength. 

Figure \ref{fig:Lsoliton} shows that, in the ``low-loss'' regime, and up to $E{s}$=5 nJ, the soliton wavelength increases with soliton energy according to Eq.~\eqref{lambdaa}. For 5<$E{s}$<15 nJ, that is, in the "low-loss" regime, the soliton wavelength still increases linearly with its energy, but with a slope that is considerably higher from the value predicted by Eq.~\eqref{lambdaa}. This is explained by the fact that Eq.~\eqref{lambdaa} describes a propagation wavelength shift. Whereas, as we have seen in Subsection \ref{subsec:FISS},  the soliton fission process suddenly produces (i.e., over a distance of a few millimeters) a spectral multiplex involving highly frequency shifted solitons. 
In the ``high-loss'' regime, i.e., for soliton energies $E{s}>$15 nJ, experimental data show an irregular distribution of Raman soliton wavelengths, oscillating around  2100 nm for the Raman 1 soliton, and 1900 nm for the Raman 2 soliton, respectively. In this regime, soliton energies may reach 50 nJ.

 \subsection{Soliton durations}
\label{subsec:PULSE}

\begin{figure}[htbp]
\centering
{\includegraphics[width=0.7\linewidth]{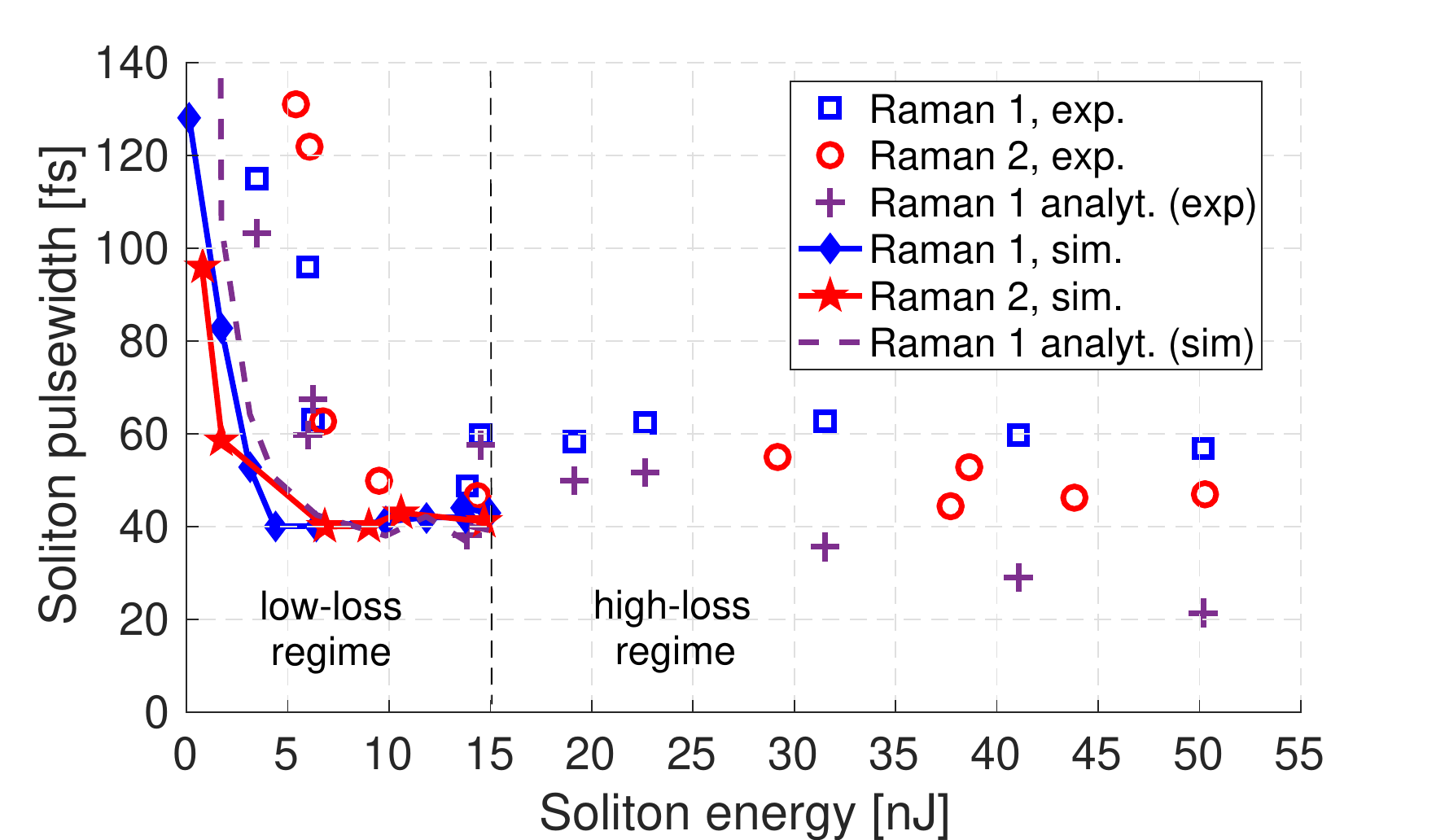}}
\caption{Experimental data (empty squares and empty circles for solitons Raman 1 and 2, respectively) and numerical simulations (solid curves) of the output soliton pulsewidth vs. soliton energy; violet crosses and violet dashed line are obtained from Eq.~\eqref{Tj} by using soliton parameters obtained from either experimental or simulation data, respectively.}
\label{fig:Tsoliton}
\end{figure}

As we shall see, the theoretical prediction of the generation of a soliton wavelength multiplex with nearly equal pulse widths is confirmed by the experimental analysis of the temporal duration of the different Raman solitons, that are generated by the fission process. Figure \ref{fig:Tsoliton} compares experimental data and numerical simulations for the pulse width of different output MM Raman solitons, as a function of their energy. Experimental pulse widths were extracted from the fit of the recorded spectra, by assuming a $\text{sech}$ shape and a transform-limited time-bandwidth product. We compared data with the analytical fundamental SM soliton relationship

\begin{equation}\label{Tj}
T_j=\frac{2\pi c|\beta_{2j}|w_j^2}{\omega_j n_2}
\end{equation}

\noindent where $\beta_{2j}$ is the dispersion at j-th soliton wavelength $\lambda_j=2\pi c/\omega_j$, and $w_j$ is the effective bullet waist of the MM soliton, which is assumed to be equal to the input waist; 
%by replacing the dispersion and wavelength of the measured output solitons; effective waist $w_j$ is assumed equal to the input waist $w_0$ for the reasons that will be illustrated below (see Figure \ref{fig:Nsoliton}); 
in Fig. \ref{fig:Tsoliton}, the violet dashed line is obtained, in the "low-loss" regime by inserting the soliton parameters as they are obtained from simulations. Whereas violet crosses are obtained by inserting in Eq.~\eqref{Tj} the soliton parameters given by the experiments. 

In the ``low-loss'' regime, for soliton energies $E_\text{s}<$15 nJ, the pulse width of individual Raman MM solitons is well approximated by the SM analytical soliton formula of Eq.~\eqref{Tj}. A train of pulses with $\text{sech}$ temporal shape, different peak powers, and wavelengths, but nearly equal pulse widths (around 45 fs) is generated for soliton energies $E_\text{s}>$5 nJ, in good agreement with numerical simulations. 

On the other hand, in the ``high-loss'' regime, that is, for $E_\text{s}>$15 nJ, the black crosses in Fig. \ref{fig:Tsoliton} show that the SM soliton formula Eq.~\eqref{Tj} would predict a narrower pulse width, when compared with respect to the experimentally observed soliton pulse durations. As a matter of fact, experimental spectra show that all MM Raman-shifted solitons converge to nearly same and constant (with respect to soliton energy) pulse width, with a value between 50 fs and 60 fs. 
%In the high loss regime, different experimental Raman solitons seem converging to a common pulsewidth (50-60 fs) that is apparently conserved for soliton energy between 10 and 50 nJ
At the same time, the wavelength shift of Raman solitons remains also clamped around a constant value (see Fig. \ref{fig:Lsoliton}).
The self-organization of a soliton multiplex with equal pulse widths and unequal amplitudes corresponds to a regime with minimal emission of dispersive waves (or radiation), since nonlinearity exactly balances the local dispersion (as it is determined by second and third-order dispersion) at each of the different Raman wavelengths \cite{KODAMA199753}.

\begin{figure}[htbp]
\centering
{\includegraphics[width=0.475\linewidth]{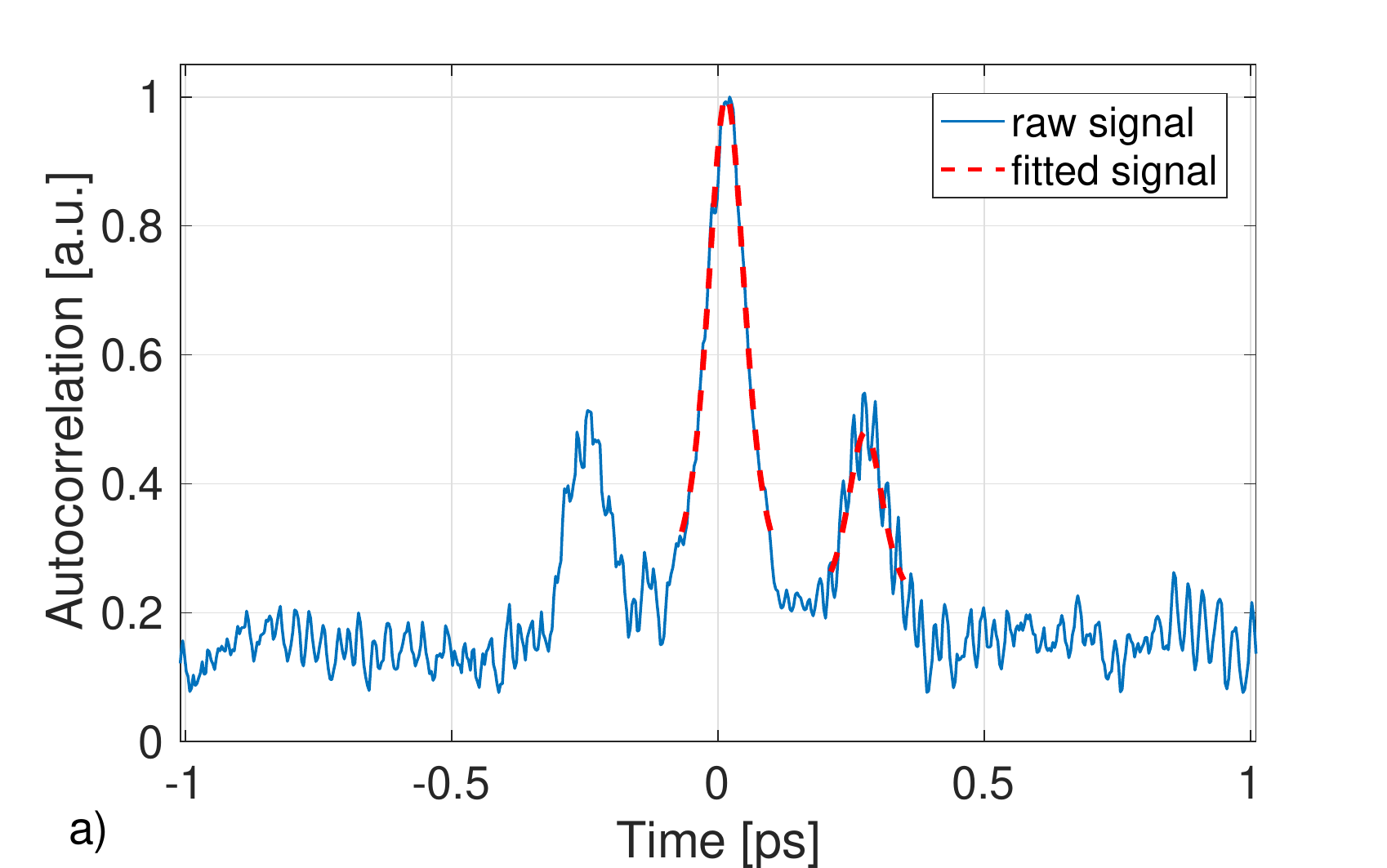}}
{\includegraphics[width=0.475\linewidth]{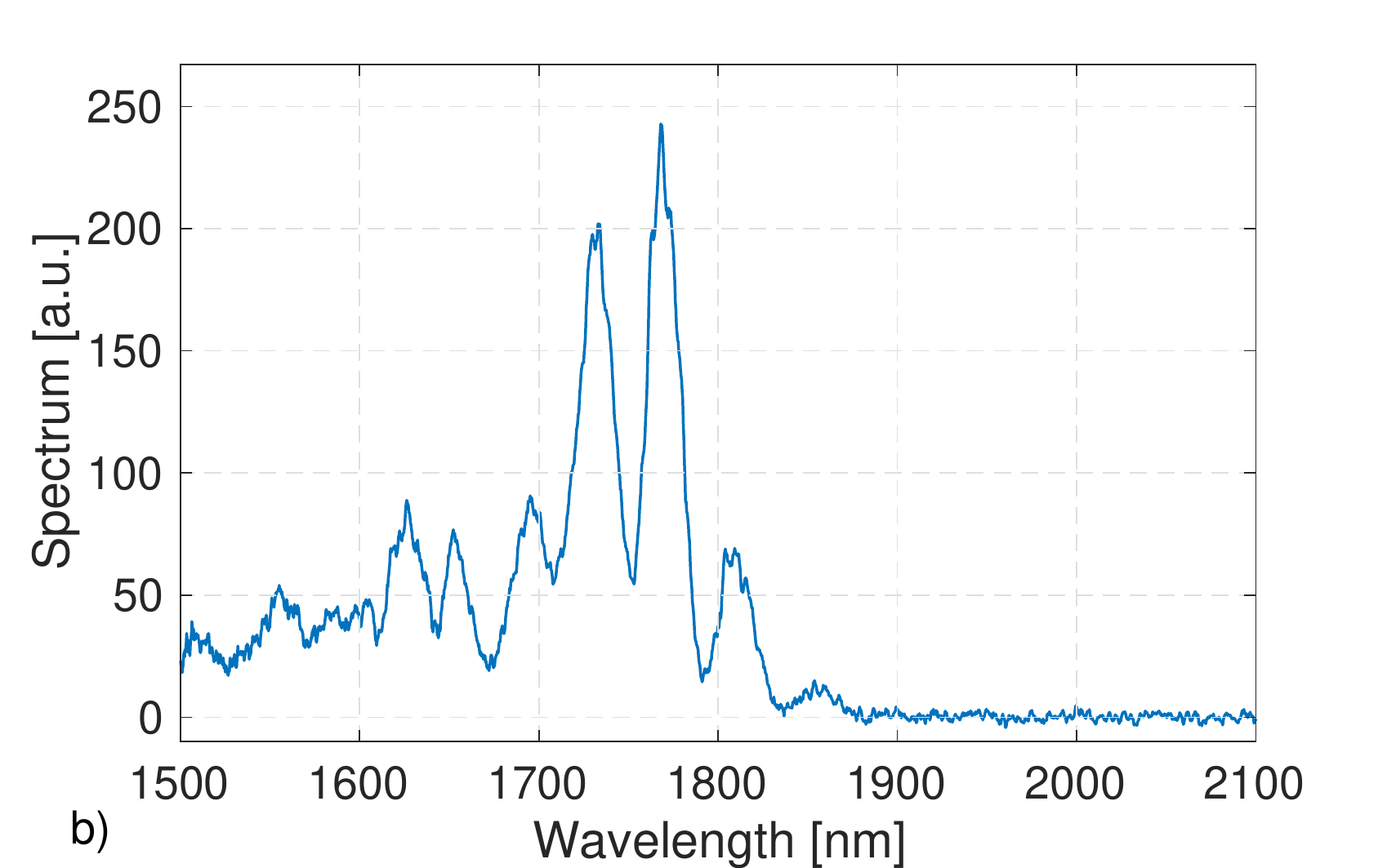}}
{\includegraphics[width=0.475\linewidth]{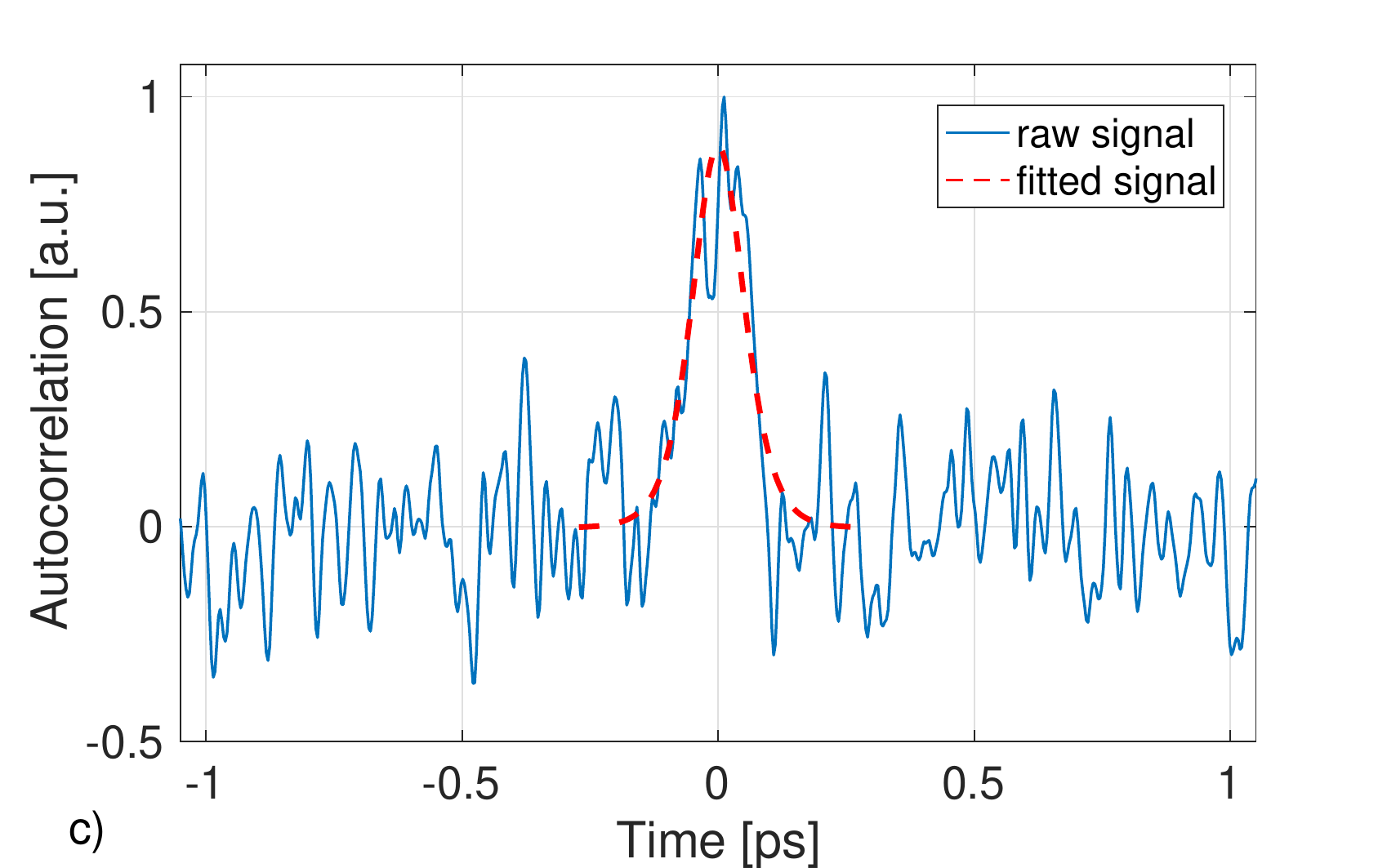}}
{\includegraphics[width=0.475\linewidth]{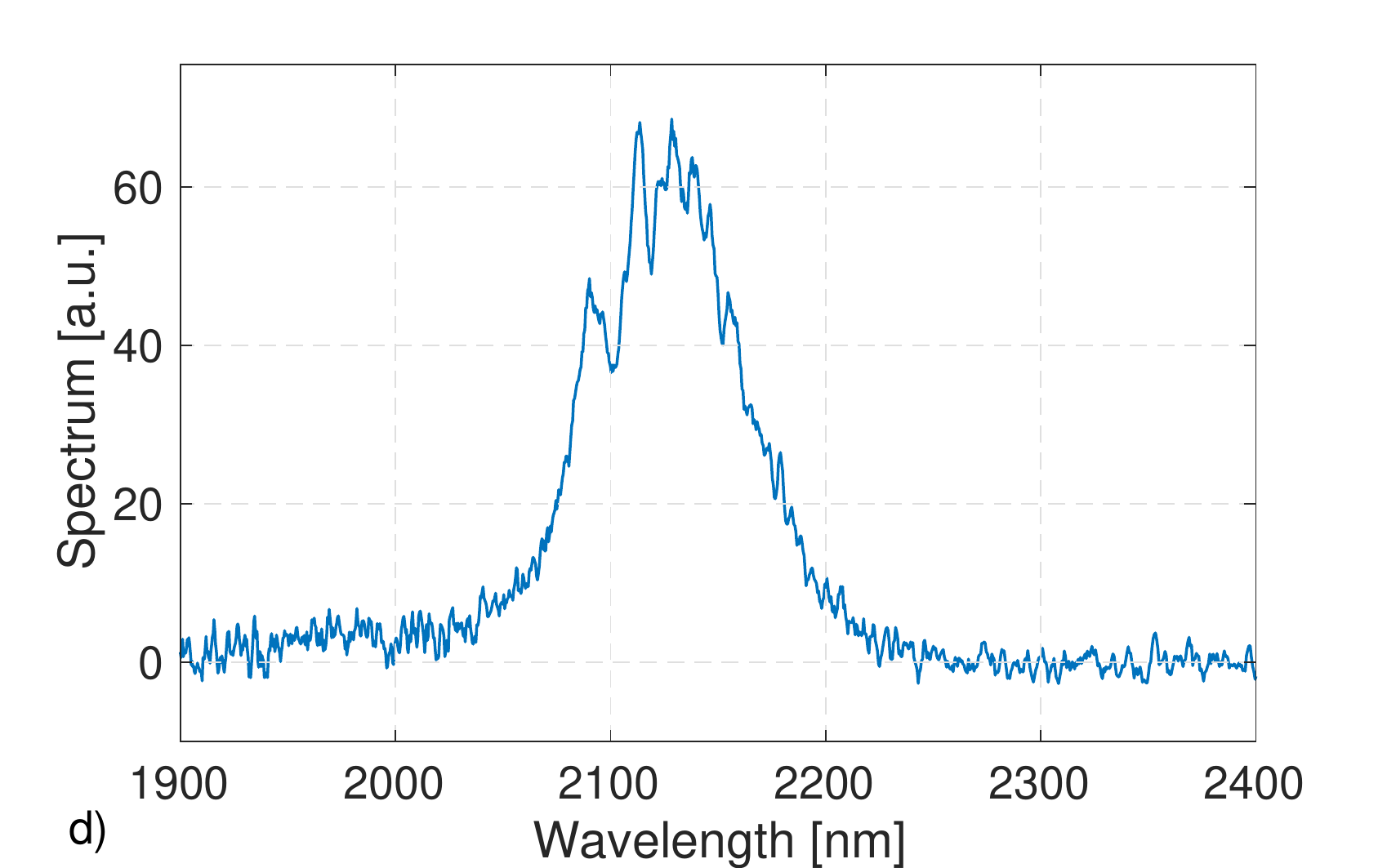}}
\caption{Experimental autocorrelation trace a) and output spectrum  b) for an input energy of 150 nJ (long-pass filtered). Autocorrelation c) and spectrum d) for an input energy of 500 nJ (band-pass filtered). Spectra are in linear scale.}
\label{fig:autocorr}
\end{figure}

In order to obtain a direct experimental confirmation of the time duration of MM solitons generated after fission, autocorrelation traces have been recorded. NIR and visible light was cut-off using filters. Fig. \ref{fig:autocorr}.(a) illustrates the case of two measured solitons, generated for an input pulse energy of 150 nJ. The solitons were isolated by using a 1500 nm long-pass filter: output solitons show a relative delay of 0.28 ps, and equal pulse widths of about 51 fs. Although crossed-pulse correlation does not provide accurate values of the individual pulse widths, obtaining cross-correlation pulses with equal widths indicates that the output solitons possess nearly the same temporal durations.
Fig. \ref{fig:autocorr}.(b) shows the corresponding measured spectrum: the interference pattern, indicating the presence of a coherent superposition of two bound solitons (soliton molecule), is clearly visible. In other cases (not shown), spectra with non-interfering patterns have been also been observed.

Fig. \ref{fig:autocorr}.(c) shows the autocorrelation of a single MM soliton, generated for 500 nJ of input energy, isolated by means of a band-pass filter centered at 2250 nm, and with 500 nm bandwidth. The Raman soliton is centered at 2110 nm, with a pulse width of 71 fs (after the filter) (see Fig. \ref{fig:autocorr}.(d)).
%Fig. 3.11a illustrates two measured solitons 
%For the input pulse energy of 150 nJ, doublets of output solitons show a relative delay of 0.52 ps, and pulsewidth of 64 and 65 fs, respectively. %Although crossed-pulse correlations does not provide accurate values of the individual pulsewidth, obtaining cross-correlation pulses with equal width indicates that output solitons possess similar pulsewidth.  Fig 3.11b shows 

 \subsection{Soliton order}
\label{subsec:ORD}

\begin{figure}[htbp]
\centering
{\includegraphics[width=0.7\linewidth]{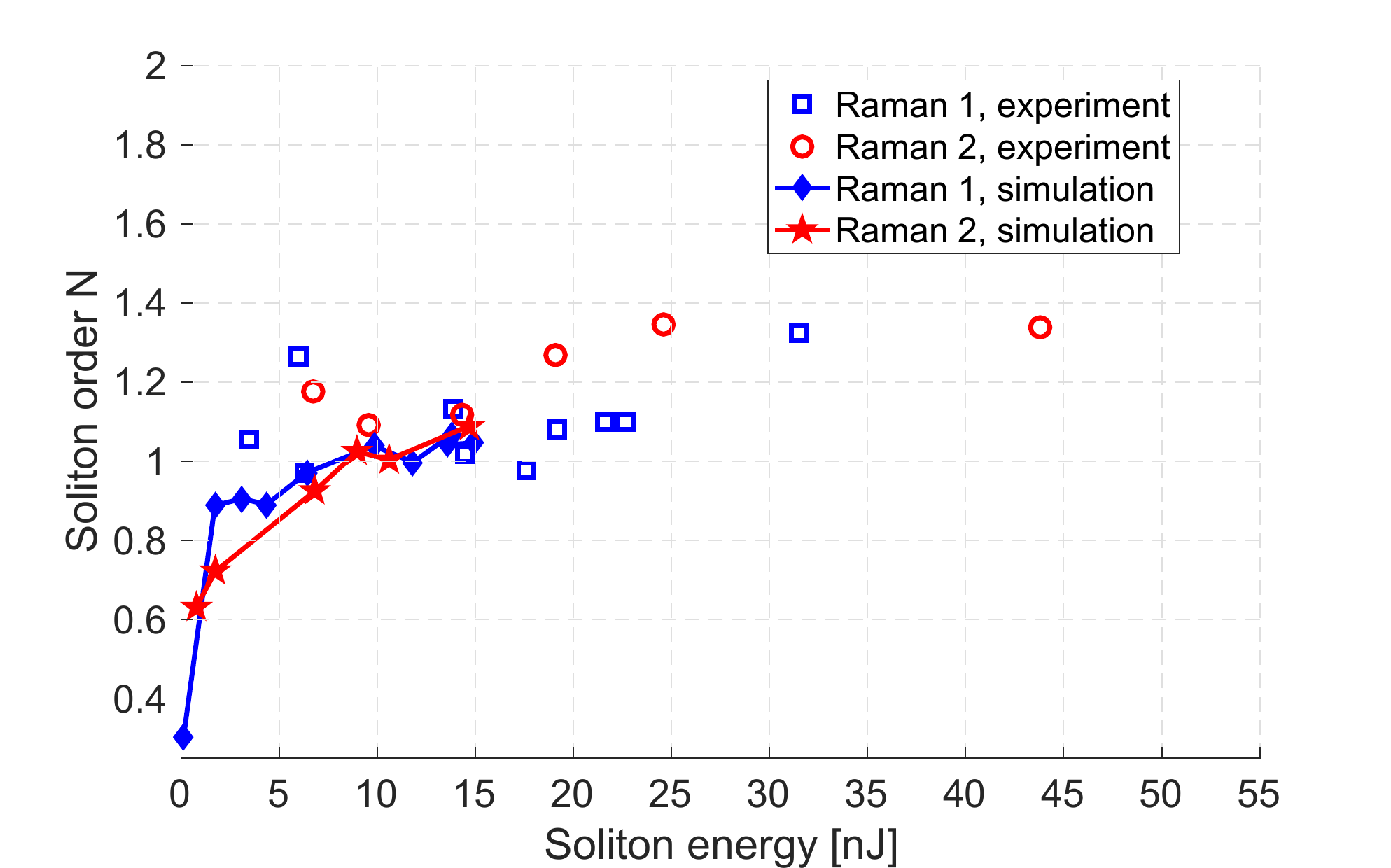}}
\caption{Experimental (dots) and numerical (solid lines) output soliton order vs. soliton energy, assuming effective soliton waist equal to the input value $w_0$.}
\label{fig:Nsoliton}
\end{figure}

In order to match experimental and numerical data with analytical results from SM soliton theory, it was necessary in Subsection \ref{subsec:PULSE} to assume that the effective bullet waist of the soliton $w_j$ is equal to the input beam waist $w_0$. This is confirmed when calculating the soliton order from the experimental and numerical data of Fig.\ref{fig:Lsoliton} and \ref{fig:Tsoliton}. For the order of the j-th fundamental MM soliton, one has

\begin{equation}\label{nj2}
N_j^2=\frac{\omega_jn_2T_{0j}E_{sj}}{2\pi c|\beta_{2j}|w_j^2}
\end{equation} 

\noindent where $E_\text{sj}$ is the $j$-th soliton energy. %nd $T_{0j}=T_{FWHMj}/1.763$. 
By using either experimental or numerical parameters, we report in Fig.\ref{fig:Nsoliton} the output soliton order as a function of its energy. Simulations in the ``low-loss'' regime, in particular, converge to $N=1$ if and only if $w_j=w_0$ is assumed for all wavelengths and energies. An attempt to use an effective soliton waist that is smaller than the input value (for example, $w_j=w_0 C^{1/4}$, where $C$ is the ratio of minimum to maximum effective area for all soliton bullets, as proposed in \cite{Ahsan:19}), leads to orders $N$ greater than unity, between 1 and 1.5.

Whereas in the ``high-loss'' regime, for $E_\text{sj}>$15 nJ, experimental soliton orders calculated with $w_j=w_0$ turn out to be in any case higher than unity, with values ranging between 1 and 1.35. This may be associated with the MM nature of the solitons, that require a higher energy with respect to SM solitons, in order to compensate for modal dispersion in addition to chromatic dispersion \cite{Wright:15,Zhu:16}.

\section{Conclusions}
We have experimentally and numerically demonstrated that high-energy, ultra-short pulse fission in a parabolic GRIN fiber produces trains of wavelength-multiplexed MM solitons with nearly equal temporal durations, and different energies. The wavelength multiplex generated during the fission process is composed by solitons with different energies, so that nearly fundamental solitons are produced at each wavelength, according to the SM soliton relationship.
%~\eqref{nj2}. 
The wavelength shift produced by the initial fission process largely dominates the propagation SSFS.
%~\eqref{lambdaa}.

For input pulse energies above 100 nJ and hundreds of kWs peak power, scattering of radiation produced during the fission process causes a substantial nonlinear energy loss. In the ``low-loss'' regime, when generated solitons possess energies below 15 nJ, the pulse width of an individual MM Raman soliton scales with its energy according to the law for fundamental SM solitons.
%eq. (\ref{Tj}). 
Conversely, in the ``high-loss'' regime, Raman solitons generated at different wavelengths have a common pulsewidth (50-60 fs). The Raman soliton time duration remains nearly constant, when its energy varies between 10 and 50 nJ. At the same time, the soliton wavelength shift appears to be clamped to a constant value.
%, in a way that is not predicted by the MM soliton theory. 
The generation of soliton wavelength multiplexes with equal pulse width and unequal amplitudes corresponds to a regime of minimal radiative wave emission, where nonlinearity balances second and third order dispersion at each soliton wavelength.

Our findings provide a new insight in the complex dynamics of MM fiber solitons, and are relevant for the development of novel fiber based sources of high-energy ultrashort pulses in the NIR and mid-infrared.

%%%%%%%%%%%%%%%%%%%%%%%%%%%%%%%%%%%%%%%%%%%%%%%%%%%%%%%%%%%%%%%%%%%%%%%%%%%%%%%%%%%
\section*{Acknowledgement}
The authors thank R. Crescenzi for technical support, and V. Couderc for measuring the GRIN fiber index profile. We acknowledge helpful discussions with D. Modotto, U. Minoni, V. Couderc, A. Tonello, and S. Murdoch.

\section*{Disclosures}

\medskip
\noindent The authors declare no conflicts of interest.
\medskip

\section*{Funding Information}
We acknowledge support from the European Research Council (ERC) under the European Union’s Horizon 2020 research and innovation program (grant No. 740355), and by the Russian Ministry of Science and Education, Grant No 14.Y26.31.0017. 

%%%%%%%%%%%%%%%%%%%%%%%%%%%%%%%%%%%%%%%%%%%%%%%%%%%%%%%%%%%%%%%%%%%%%%%%%%%%%%%%%%%

% Bibliography
\bibliography{BIBLIO_MultimodeTemp3}

\end{document}